\documentclass[a4paper,UKenglish,cleveref]{lipics-v2021}
\nolinenumbers

\title{Verification and Evaluation of A Comprehensive Quantum Simulation Compiler}                      

\author{Bhuwan Joshi}{Iowa State University, Iowa, USA}{bbjoshi@iastate.edu}{}{}

\author{Zoukarneini Tagba}{Iowa State University, Iowa, USA}{difaizi@iastate.edu}{}{}

\author{Zhi Xiang Chong}{Iowa State University, Iowa, USA}{ianchong@iastate.edu}{}{}

\author{Liyi Li}{Iowa State University, Iowa, USA}{liyili2@iastate.edu}{}{}

\authorrunning{B. Joshi and Z. Tagba and Z. Chong and L. Li} 

\Copyright{Bhuwan Joshi and Zoukarneini Tagba and Zhi Xiang Chong and Liyi Li}



\usepackage{stmaryrd}
\usepackage{booktabs}   \usepackage{subcaption} \usepackage{bussproofs}
\usepackage[cal=boondoxo]{mathalfa}
\DeclareMathAlphabet{\mathpzc}{OT1}{pzc}{m}{it}
\usepackage{color}
\usepackage{xspace}

\usepackage{listings}
\newcommand{\ignore}[1]{{}}

\newcommand{\syntaxDef}[3]{\rulebox{\syntaxKeyword$#1\mathrel{::=}{#2}$ \ifthenelse{\equal{#3}{}}{}{[#3]}}}

\makeatletter
\newcommand{\shorteq}{\settowidth{\@tempdima}{-}\resizebox{\@tempdima}{\height}{=}}
\makeatother

\usepackage{xcolor}
\usepackage{colortbl}

\usepackage[utf8]{inputenc}

\usepackage{amsmath}
\usepackage{hyperref}

\usepackage{subcaption}
\usepackage{wrapfig}

\usepackage{xspace}
\usepackage{float}
\usepackage{balance}

\usepackage{textcomp} 

\usepackage{changepage}
\usepackage{array,etoolbox}

\usepackage{algorithm}
\usepackage{algorithmic}

\preto\tabular{\setcounter{magicrownumbers}{0}}
\newcounter{magicrownumbers}

\usepackage{cleveref} 

\usepackage{wrapfig}
\usepackage{caption}

\usepackage{longtable}
\usepackage{tabu}

\usepackage{wasysym}
\usepackage{mathpartir}

\usepackage{mathtools} \usepackage{xfrac}

\usepackage[qm,braket]{qcircuit}

\newcommand{\qif}[2]{\texttt{{if}\,(}{#1}\texttt{)\,\{}{#2}\texttt{\}}}

\colorlet{kwd}{black!80!green}
\definecolor{spec1}{RGB}{78, 131, 162}
\definecolor{spec0}{RGB}{66, 102, 136}
\definecolor{lespec}{RGB}{30, 80, 180}
\colorlet{spec}{lespec}
\colorlet{auto}{lespec!35!lightgray}
\colorlet{stack}{magenta}

\newcommand{\sqir}{SQIR\xspace}

\newcommand{\qwire}{{Qwire}\xspace}
\newcommand{\qbricks}{{QBricks}\xspace}

\newcommand{\voqc}{\textsc{voqc}\xspace}

\newcommand{\myparagraph}[1]{\noindent\paragraph{\textbf{#1}}}

\usepackage{tikz}

\newcommand{\uapp}[2]{#1\, #2}

\newcommand{\slen}[1]{|#1|}
\newcommand{\pau}{\mathpzc{P}}

\newcommand{\CC}{\ensuremath{\mathbb{C}}\xspace}

\usepackage{pgf}

\usepackage{tikz} \usetikzlibrary{arrows,shapes.misc,shapes.geometric, shapes.arrows,shapes.callouts,
  shapes.gates.logic.US,
  chains,matrix,positioning,scopes,decorations.pathmorphing,decorations.text,
  decorations.pathreplacing, shadows,automata,
  fit, calc, arrows.meta
}

\tikzset{ machine/.style={
rectangle,
minimum width=25mm,
    minimum height=18mm,
    text width=24mm,
align=center,
very thick,
    draw=black,
color=black,
    fill=white,
}
}

\DeclarePairedDelimiter\abs{\lvert}{\rvert}
\DeclarePairedDelimiter\norm{\lVert}{\rVert}

\makeatletter
\let\oldabs\abs
\def\abs{\@ifstar{\oldabs}{\oldabs*}}
\let\oldnorm\norm
\def\norm{\@ifstar{\oldnorm}{\oldnorm*}}
\makeatother

\makeatletter
\DeclareRobustCommand{\vardivision}{\mathbin{\mathpalette\@vardivision\relax}}
\newcommand{\@vardivision}[2]{\reflectbox{$\m@th\smallsetminus$}}
\makeatother

\usepackage{booktabs}

\usepackage{alltt}
\usepackage{listings,lstcoq}
\definecolor{ltblue}{rgb}{0,0.4,0.4}
\definecolor{dkblue}{rgb}{0,0.1,0.6}
\definecolor{dkgreen}{rgb}{0,0.35,0}
\definecolor{dkviolet}{rgb}{0.3,0,0.5}
\definecolor{dkred}{rgb}{0.5,0,0}
\usepackage[export]{adjustbox}

\newcommand{\code}[1]{{\small\texttt{#1}}}

\DeclareMathOperator*{\Motimes}{\text{\raisebox{0.25ex}{\scalebox{0.8}{$\bigotimes$}}}}

\newcommand{\sapp}[2]{#1\circ #2}

\newcommand{\zero}{\mathpzc{O}}

\newcommand{\sdag}[1]{#1^{\dag}}

\newcommand{\rocq}{\textsc{Rocq}\xspace}

\newcommand{\sminus}{\texttt{-}}
\newcommand{\splus}{\texttt{+}}
\newcommand{\eexp}[1]{\cn{exp}(\sminus i #1)}
\newcommand{\pexp}[1]{\cn{exp}(i #1)}

\newcommand{\iseq}[2]{{#1}\,{#2}}

\newcommand{\qsndl}{$\textsc{QBlue}_{\cn{BE}}$\xspace}
\newcommand{\qsnd}{$\textsc{QBlue}$\xspace}
\newcommand{\qblue}{\qsnd}
\newcommand{\Hilb}{\mathcal H}

\newcommand{\cn}[1]{\texttt{#1}}

\newcommand{\denote}[1]{\llbracket #1 \rrbracket\xspace}
\newcommand{\dabs}[1]{|\!| #1 |\!|}

\newcommand{\lift}[1]{\ensuremath{\lceil #1 \rceil}\xspace}

\usepackage{newunicodechar}
\let\Alpha=A
\let\Beta=B
\let\Epsilon=E
\let\Zeta=Z
\let\Eta=H
\let\Iota=I
\let\Kappa=K
\let\Mu=M
\let\Nu=N
\let\Omicron=O
\let\omicron=o
\let\Rho=P
\let\Tau=T
\let\Chi=X

\newunicodechar{Α}{\ensuremath{\Alpha}}
\newunicodechar{α}{\ensuremath{\alpha}}
\newunicodechar{Β}{\ensuremath{\Beta}}
\newunicodechar{β}{\ensuremath{\beta}}
\newunicodechar{Γ}{\ensuremath{\Gamma}}
\newunicodechar{γ}{\ensuremath{\gamma}}
\newunicodechar{Δ}{\ensuremath{\Delta}}
\newunicodechar{δ}{\ensuremath{\delta}}
\newunicodechar{Ε}{\ensuremath{\Epsilon}}
\newunicodechar{ε}{\ensuremath{\epsilon}}
\newunicodechar{ϵ}{\ensuremath{\varepsilon}}
\newunicodechar{Ζ}{\ensuremath{\Zeta}}
\newunicodechar{ζ}{\ensuremath{\zeta}}
\newunicodechar{Η}{\ensuremath{\Eta}}
\newunicodechar{η}{\ensuremath{\eta}}
\newunicodechar{Θ}{\ensuremath{\Theta}}
\newunicodechar{θ}{\ensuremath{\theta}}
\newunicodechar{ϑ}{\ensuremath{\vartheta}}
\newunicodechar{Ι}{\ensuremath{\Iota}}
\newunicodechar{ι}{\ensuremath{\iota}}
\newunicodechar{Κ}{\ensuremath{\Kappa}}
\newunicodechar{κ}{\ensuremath{\kappa}}
\newunicodechar{Λ}{\ensuremath{\Lambda}}
\newunicodechar{λ}{\ensuremath{\lambda}}
\newunicodechar{Μ}{\ensuremath{\Mu}}
\newunicodechar{μ}{\ensuremath{\mu}}
\newunicodechar{Ν}{\ensuremath{\Nu}}
\newunicodechar{ν}{\ensuremath{\nu}}
\newunicodechar{Ξ}{\ensuremath{\Xi}}
\newunicodechar{ξ}{\ensuremath{\xi}}
\newunicodechar{Ο}{\ensuremath{\Omicron}}
\newunicodechar{ο}{\ensuremath{\omicron}}
\newunicodechar{Π}{\ensuremath{\Pi}}
\newunicodechar{π}{\ensuremath{\pi}}
\newunicodechar{ϖ}{\ensuremath{\varpi}}
\newunicodechar{Ρ}{\ensuremath{\Rho}}
\newunicodechar{ρ}{\ensuremath{\rho}}
\newunicodechar{ϱ}{\ensuremath{\varrho}}
\newunicodechar{Σ}{\ensuremath{\Sigma}}
\newunicodechar{σ}{\ensuremath{\sigma}}
\newunicodechar{ς}{\ensuremath{\varsigma}}
\newunicodechar{Τ}{\ensuremath{\Tau}}
\newunicodechar{τ}{\ensuremath{\tau}}
\newunicodechar{Υ}{\ensuremath{\Upsilon}}
\newunicodechar{υ}{\ensuremath{\upsilon}}
\newunicodechar{Φ}{\ensuremath{\Phi}}
\newunicodechar{φ}{\ensuremath{\phi}}
\newunicodechar{ϕ}{\ensuremath{\varphi}}
\newunicodechar{Χ}{\ensuremath{\Chi}}
\newunicodechar{χ}{\ensuremath{\chi}}
\newunicodechar{Ψ}{\ensuremath{\Psi}}
\newunicodechar{ψ}{\ensuremath{\psi}}
\newunicodechar{Ω}{\ensuremath{\Omega}}
\newunicodechar{ω}{\ensuremath{\omega}}

\newunicodechar{ℕ}{\ensuremath{\mathbb{N}}}
\newunicodechar{∅}{\ensuremath{\emptyset}}

\newunicodechar{∙}{\ensuremath{\bullet}}
\newunicodechar{≈}{\ensuremath{\approx}}
\newunicodechar{≅}{\ensuremath{\cong}}
\newunicodechar{≡}{\ensuremath{\equiv}}
\newunicodechar{≤}{\ensuremath{\le}}
\newunicodechar{≥}{\ensuremath{\ge}}
\newunicodechar{≠}{\ensuremath{\neq}}
\newunicodechar{∀}{\ensuremath{\forall}}
\newunicodechar{∃}{\ensuremath{\exists}}
\newunicodechar{±}{\ensuremath{\pm}}
\newunicodechar{∓}{\ensuremath{\pm}}
\newunicodechar{·}{\ensuremath{\cdot}}
\newunicodechar{⋯}{\ensuremath{\cdots}}
\newunicodechar{…}{\ensuremath{\ldots}}
\newunicodechar{∷}{~\mathrel{:\!\!\!:}~}
\newunicodechar{×}{\ensuremath{\times}}
\newunicodechar{∞}{\ensuremath{\infty}}
\newunicodechar{→}{\ensuremath{\to}}
\newunicodechar{←}{\ensuremath{\leftarrow}}
\newunicodechar{⇒}{\ensuremath{\Rightarrow}}
\newunicodechar{↦}{\ensuremath{\mapsto}}
\newunicodechar{↝}{\ensuremath{\leadsto}}
\newunicodechar{∨}{\ensuremath{\vee}}
\newunicodechar{∧}{\ensuremath{\wedge}}
\newunicodechar{⊢}{\ensuremath{\vdash}}
\newunicodechar{⊣}{\ensuremath{\dashv}}
\newunicodechar{∣}{\ensuremath{\mid}}
\newunicodechar{∈}{\ensuremath{\in}}
\newunicodechar{⊆}{\ensuremath{\subseteq}}
\newunicodechar{⊂}{\ensuremath{\subset}}
\newunicodechar{∪}{\ensuremath{\cup}}
\newunicodechar{⋓}{\ensuremath{\Cup}}
\newunicodechar{∉}{\ensuremath{\not\in}}
\newunicodechar{√}{\ensuremath{\sqrt}}

\newunicodechar{⊸}{\ensuremath{\multimap}}
\newunicodechar{⊗}{\ensuremath{\otimes}}
\newunicodechar{⨂}{\ensuremath{\bigotimes}}
\newunicodechar{⊕}{\ensuremath{\oplus}}
\newunicodechar{〈}{\ensuremath{\langle}}
\newunicodechar{⟨}{\ensuremath{\langle}}
\newunicodechar{⟩}{\ensuremath{\rangle}}
\newunicodechar{〉}{\ensuremath{\rangle}}
\newunicodechar{¡}{\ensuremath{\upsidedownbang}}
\newunicodechar{∘}{\ensuremath{\circ}}
\newunicodechar{†}{\ensuremath{\dagger}}
\newunicodechar{⊤}{\ensuremath{\top}}
\newunicodechar{⊥}{\ensuremath{\bot}}

\newunicodechar{〚}{\ensuremath{\llbracket}}
\newunicodechar{〛}{\ensuremath{\rrbracket}}

 \usepackage{bbold}
\usepackage{pifont}\usepackage{multicol,tabularx,capt-of}
\usepackage{multirow}

\begin{document}

\maketitle

\begin{abstract}
 Hamiltonian simulation is a central application of quantum computing, yet most existing compilers offer no machine-checked guarantee that the circuits they produce approximate the intended evolution.  The previous QBlue compiler verifies a Hamiltonian simulation compiler containing only Trotter-based algorithms. It takes a Hamiltonian written over Pauli strings, a simulation time, and an error bound, and produces a circuit or analog schedule that is proven, in the Rocq proof assistant, to approximate the ideal time evolution within that bound and proves a single end-to-end error bound that composes the errors of the individual stages; the compiler uses it to select, by resource estimation, the cheapest algorithm that meets the user's bound.  This paper advances the verification efforts with two further steps. First, we extend QBlue to \qsndl by verifying another important compilation method for Hamiltonian simulation: block-encoding-based (BE-based) algorithms. Second, we empirically evaluate 279 Hamiltonian simulation programs against the state-of-the-art BE-based compiler, PennyLane. For many programs, our compiler shows a $94\%$ gate reduction compared to PennyLane. Additionally, we also discover issues and bugs in PennyLane. We also compare Trotter- and BE-based algorithms, providing a practical comparison for the long-standing debate over which algorithms perform better. Our results surprisingly show that Trotter-based algorithms may gain a practical advantage, even though BE algorithms have a theoretical advantage.

 \end{abstract}

\section{Introduction}\label{sec:intros}

Quantum simulation is a flagship application of quantum computing, promising to model physical systems~\cite{Zhang_2021,osti_1782924,du2023multinucleon}, chemical reactions~\cite{Lanyon2010,Cao2019,Evangelista2023,RevModPhys.92.015003}, and materials~\cite{Emani2021,Fedorov2021,wcms.1481} with precision beyond classical capabilities~\cite{feynman1982simulating}.  
In Quantum simulation, a physical system is described by a Hamiltonian $\hat{H}$, and simulating it for time $r$ means implementing the unitary $\exp(-ir\hat{H})$ as a circuit or an analog schedule that a device can run.  
The algorithms used in practice~\cite{cao2024marqsimreconcilingdeterminismrandomness,Andrew2018,Campbell_2019,Low_2019,Berry_2015,Low_2017,Martyn_2021,Andrew2018} only approximate this unitary, so correctness is quantitative: a compiled program is correct when the unitary it denotes is within a stated distance of $\exp(-ir\hat{H})$.  The value of the answer then depends on how large the error is and on whether the compiler knows it.  
Current compilers for Hamiltonian simulation~\cite{10.1145/3470496.3527394,10.1145/3632923,cao2024marqsimreconcilingdeterminismrandomness,10.1145/3503222.3507715} optimize aggressively, but none offers a machine-checked guarantee that its output is close to the intended evolution.  
A miscompiled simulation fails silently: the circuit runs, returns numbers, and the numbers describe a different physical system.  
Choosing the algorithm is a second difficulty.  Many have been proposed, each claiming superior performance, and the cheapest option depends on the Hamiltonian, the error tolerance, and the platform.

Previously, \qsnd \cite{li2025verifiedcompilerquantumsimulation} develops a verified quantum simulation compiler for Trotter-based Hamiltonian simulation compilation, based on the Trotter-Suzuki decomposition.
\qsnd takes as input a Hamiltonian expression built from Pauli strings, a simulation time~$r$, and an error bound $\epsilon$.  This input format is general: the Pauli strings on $n$ qubits form a basis for all operators on $n$ qubits, so every qubit Hamiltonian can be written this way. 
The compiler implements four different Trotter-based algorithms, namely first-order and second-order Trotterization~\cite{529425}, QDrift~\cite{Campbell_2019}, and MarQSim~\cite{cao2024marqsimreconcilingdeterminismrandomness}, which turn $\exp(-ir\hat{H})$ into a sequence of exponentials of single Pauli strings, and synthesizes that sequence for one of three backends: standard digital gate sequences, IBM analog superconducting hardware~\cite{10.1145/3632923}, and a trapped-ion processor whose global drives implement multi-qubit interactions directly~\cite{Kyprianidis_2024,Richerme_2025}.
The compiler's main result is a correctness guarantee, with mechanized verification in the \rocq proof assistant~\cite{bertot2013interactive}. 
The guarantee states that for a Hermitian Pauli-string Hamiltonian $\hat{H}$, a time~$r$, and an error bound $\epsilon$, the compiled program $U$ satisfies $\lVert \llbracket U \rrbracket(\psi) - \llbracket \exp(-ir\hat{H}) \rrbracket(\psi) \rVert < \epsilon$ for every normalized state $\psi$.  To meet $\epsilon$, the compiler computes, symbolically from the structure of~$\hat{H}$, an upper bound on the error of each candidate algorithm on the target backend, and sets the algorithm's parameters, such as the number of Trotter steps, to ensure that the error bound stays below $\epsilon$.  Among the algorithms that meet $\epsilon$, it picks the one with the smallest estimated cost.

\begin{figure}[t]
\includegraphics[width=0.9\textwidth]{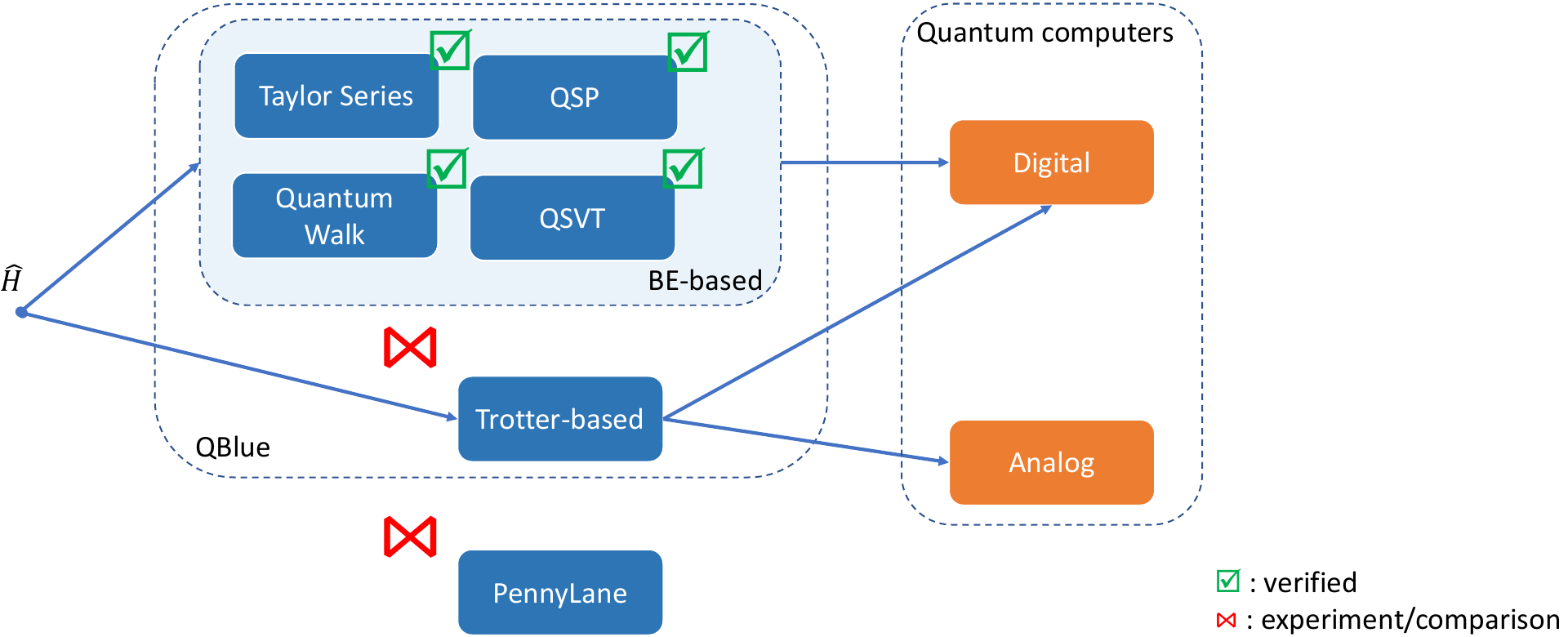}
\caption{QBlue project flow; the top is \qsndl; we compare it against PennyLane and compare BE-based algorithms against Trotter-based ones in QBlue.}
\label{fig:projectflow}
\end{figure}

\qsnd is solely Trotter-based and ignores other Hamiltonian simulation algorithms, i.e., BE-based algorithms.
Instead of performing Trotter-Suzuki decomposition, BE-based algorithms recognize that the $\exp(-ir\hat{H})$ implementation can be expanded via expansion formulas, such as Taylor and Jacobi-Anger, yielding an expression being a linear sum of Pauli strings, which can be implemented on a quantum computer via the block encoding (BE) technique --- it mainly connects linear sum operations with quantum controlled gates.

In this paper, we extend \qblue with BE-based Hamiltonian simulation algorithms and conduct an empirical study comparing different compilation algorithms, as shown in \Cref{fig:compilationprocess}.
Our first contribution is to identify the compilation pipeline of BE-based algorithms, \qsndl, and verify all these algorithms with correctness guarantees, established in \rocq.
During verification, we discover issues and bugs in the existing platform, PennyLane \cite{bergholm2022pennylaneautomaticdifferentiationhybrid}; we do this by evaluating on large benchmarks and comparing \qsndl compilation with PennyLane to identify these issues and bugs.
To our knowledge, it is the first verified compiler for Hamiltonian simulation that carries a mechanized error bound across several BE-based algorithms.

Another contribution is an empirical study comparing Trotter- and BE-based compilation algorithms.
BE-based algorithms are expected to have better theoretical complexity than Trotter-based algorithms \cite{Andrew2018}, but their practical advantage is unclear.
In this paper, we compile these algorithms using a comprehensive benchmark of real-world quantum chemistry and physics programs.
We examine the qubit and gate counts of the compiled circuits for different algorithms to understand the advantage of the two different groups.
We uncover surprising results. Although BE-based algorithms offer a theoretical advantage in gate savings for longer simulation times and tighter error bounds, for real-world cases with reasonable simulation time and error bounds, Trotter-based algorithms offer a practical advantage in qubit and gate counts.

\myparagraph{Summary of Contributions}
We develop an end-to-end verification of a BE-based Hamiltonian simulation compiler, mechanizing each pipeline step in \rocq with a verified theorem (\Cref{fig:compilationprocess}).
We also conduct an empirical study of BE-based Hamiltonian simulation compilers.
In the study, we evaluate \qsndl on 279 Hamiltonian simulation programs, from spin models to molecular Hamiltonians, and compare it against an unverified state-of-the-art compiler, PennyLane~\cite{bergholm2022pennylaneautomaticdifferentiationhybrid}, as well as examine aspects of a Hamiltonian simulation compiler, including comparing Trotter-based and BE-based algorithms and comparing algorithms within the BE group.
One key difference is that Trotter-based algorithms suit analog quantum computers, while BE-based algorithms are implementable only on digital quantum computers.
We compare the efficiency of Trotter-based algorithms on an analog quantum computer with BE-based algorithms on a digital quantum computer and find that the former is more efficient.
We summarize the paper's contributions as follows.

\begin{itemize}
  \item A verified compilation pipeline that spans constraint, dynamic, and the compiler targeting BE-based algorithms with certified error bounds;
  \item A mechanized and end-to-end compilation theorem. We develop a library for every step in the \qsndl Hamiltonian simulation compiler, serving as the standard for BE-based Hamiltonian simulation;
  \item A comprehensive evaluation of Hamiltonian simulation compilation based on BE-based algorithms, comparing the \qsndl compiler with the industry compiler, PennyLane. We discover issues in PennyLane and show that our compiler saves more gates than PennyLane, i.e., we save more than $90\%$ of the gates in many cases;
  \item An empirical study comparing BE- and Trotter-based algorithms, including comparisons on digital and analog platforms, as well as an examination of different BE-based algorithms;
  \item A solid benchmark of 279 programs, including small, medium, and large real-world programs, where these benchmarks can serve as a standard set for Hamiltonian simulations.
\end{itemize}

\myparagraph{Paper Roadmap}
The remainder of the paper is organized as
follows. \Cref{sec:overview} walks through \qsndl on an Ising model and introduces the background along the way.
\Cref{sec:compilation} describes the compilation pipeline and proves its correctness.  \Cref{sec:arith-oqasm,sec:discussion} evaluate
\qsndl, \Cref{sec:related} discusses related work, and
\Cref{sec:conclusion} concludes.

\section{Background and Overview}
\label{sec:overview}

This section follows a small Hamiltonian through \qsndl, from its physical description to a circuit, and introduces the background each step needs.  \Cref{fig:compilationprocess} shows the stages.

\subsection{Background about Quantum State}
\label{sec:background}

We first provide background information on quantum states used in quantum simulation.

\begin{figure}[t]
{\small
  \[
  \begin{array}{c}
  {\hspace{-1em}
  \begin{array}{l@{\;\;\;}l@{\;}c@{\;}l @{\qquad} l@{\;\;\;}l@{\;}c@{\;}l @{\qquad} l@{\;\;\;}l@{\;}c@{\;}l @{\qquad} l@{\;\;\;}l@{\;\;}l @{\qquad} l@{\;\;\;}l@{\;\;}l @{\qquad} l@{\;\;\;}l@{\;\;}l}
  \text{Nat. Num} & m, n, j, k,g & \in & \mathbb{N}
&
\text{Reals} & \theta, r & \in & {\mathbb{R}}
&
\text{Complex} & z & \in & {\mathbb{C}}
\\[0.2em]
 \text{Variable} & x,y,f &&&
 
\text{Bit/Bool} & b & ::= & 0 \mid 1
& \text{Basis Vector} & \eta & \in & \ket{b}
\end{array}
}
\\[0.5em]
  \begin{array}{l@{\quad}l@{\;\;}c@{\;\;}l@{\qquad\qquad} l@{\quad}l@{\;\;}c@{\;\;}l@{}c@{\;\;\;}l} 
n\text{-Site Basis-ket State} & w & ::= &  {\displaystyle \Motimes_{k=0}^{n-1} \eta_{k} }
&
\text{Quantum State} & \psi & ::= & {\displaystyle {\sum_j} z_j w_j  } &\mid& \zero
    \end{array}
        \end{array}
  \]
}
  \caption{\qsnd state structure for particles. We abbreviate $z \cdot \Motimes_{k=0}^{0} \eta_{k} \equiv z\, \eta_0$ and $z_1\eta_1 \otimes z_2\eta_2\equiv (z_1 * z_2)\eta_1 \eta_2$.}
  \label{fig:data}
\end{figure}

In quantum computing, a system's state is described by a unit vector in a Hilbert space. 
A quantum state consists of one or more \emph{qubits} (\emph{sites}, physical locations), each of which is a two-dimensional Hilbert space $\Hilb_2$.
When modeling particle systems via Hamiltonians, a single qubit in a quantum state is typically called a \emph{site}, while a quantum-computation state is typically called a \emph{qubit}. They refer to similar concepts, but the former describes a unit vector in a particle system, while the latter describes quantum computation. In this paper, we use 'sites' generically and 'qubits' to refer to the transformed Pauli-based Hamiltonians and quantum unitary operations.

A single qubit can be represented as a two-dimensional vector $\begin{psmallmatrix} z_1 \\ z_2 \end{psmallmatrix}$, where $z_1$ and $z_2$ are complex amplitudes with $|z_1|^2 + |z_2|^2 = 1$. Using Dirac notation, this is written as $z_1\ket{0} + z_2\ket{1}$, with $\ket{0}$ and $\ket{1}$ as the computational basis states. When both $z_1$ and $z_2$ are non-zero, the qubit is in a \emph{superposition} of $\ket{0}$ and $\ket{1}$. Multi-qubit data is constructed via the tensor product, e.g., $\ket{0}\otimes\ket{1} = \ket{01}$. However, not all multi-qubit states can be written as a tensor product; some are entangled, such as the Bell pair $\frac{1}{\sqrt{2}}(\ket{00}+\ket{11})$.

\Cref{fig:data} defines the quantum-state syntax used in this paper, formalizing such systems using \emph{computational basis vectors and basis kets}. A single-site basis vector is written as the ket $\eta_k = \ket{b}$, where $k \in [0, 1]$ identifies one of the local basis vectors. A composite $n$-site basis-ket state is then expressed as $w = \Motimes_{k=0}^{n-1} \eta_k$. For example, two sites in basis vectors $\ket{0}$ and $\ket{1}$ compose into the joint state $\ket{0} \otimes \ket{1}$, which we abbreviate as $\ket{0}\ket{1}$. 

A single-site basis vector can be scaled by a complex amplitude $z$, forming a basis-ket $z \ket{k}$. For an $n$-site system, we write basis-kets as $w = z\cdot\Motimes_{k=0}^{n-1} \eta_k$, and associate amplitudes component-wise. For example, the state $z_1 \eta_1 \otimes z_2 \eta_2$ can be abbreviated as $(z_1 * z_2) \eta_1 \eta_2$. In general, a quantum superposition state is written canonically as $\psi = \sum_j z_j w_j$. We also include the zero vector $\zero$ to represent the null state, which plays a role in the semantics of second quantization.
As a convenient notation, we denote $\ket{j}_n$ to refer to an $n$ qubit system, and it is equal to $\ket{j[0]}\otimes \ket{j[1]} \otimes ... \otimes \ket{j[n-1]}$.

\subsection{The Example: A Two-Site Hubbard Model}
\label{sec:hubbard-example}

The Hubbard model~\cite{ec556129} describes electrons that hop between
neighboring sites of a lattice and repel one another.  We use its
smallest instance~\cite{melo_2021}: two sites, labeled $0$ and $1$,
holding spinless fermions, so that each site is either empty or holds
one particle.  A state of the system is a unit vector in
$\mathbb{C}^2 \otimes \mathbb{C}^2$ with basis states
$\ket{m}\ket{n}$, where $m, n \in \{0,1\}$ record the occupation of
sites $0$ and $1$.  A general state is a superposition
$\sum_{m,n} z_{mn}\ket{m}\ket{n}$ with $\sum_{m,n} |z_{mn}|^2 = 1$.
This is exactly the state space of two qubits, which is why such
models are natural targets for a quantum computer.

To describe the model, people use a Hamiltonian, syntax in Appendix \ref{sec:formal}, to describe the physical constraints.
In this model, we sum over different constraints, and the Hubbard model has two main components: one describes electron hopping between sites $1$ and $0$, and the other describes the repulsion between the two sites.
The Hamiltonian written in the Pauli-string form is given as follows, where the .$-\tfrac12\, X(0)\circ X(1) - \tfrac12\, Y(0)\circ Y(1)$ term is the movement term and the $\tfrac12\, Z(0)\circ Z(1)$ term is the repulsion. 
\[
\hat{H} = -\tfrac12\, X(0)\circ X(1) - \tfrac12\, Y(0)\circ Y(1)
          + \tfrac12\, Z(0)\circ Z(1).
\tag{2.3}\label{eq:ex}
\]

As in the Hubbard model example above, a Hamiltonian is a Hermitian matrix, $\hat{H} = \hat{H}^\dagger$, whose eigenvalues are the energy levels of the system.  Simulating the system for time $r$ means applying the unitary $\exp(-ir\hat{H})$, which is unitary precisely because $\hat{H}$ is Hermitian.  This unitary is what \qsnd compiles.  

The example above uses Pauli matrix operations. A quantum computer acts on qubits through the Pauli matrices ($\pau$)
\[
I = \begin{psmallmatrix}1&0\\0&1\end{psmallmatrix},\quad
X = \begin{psmallmatrix}0&1\\1&0\end{psmallmatrix},\quad
Y = \begin{psmallmatrix}0&-i\\i&0\end{psmallmatrix},\quad
Z = \begin{psmallmatrix}1&0\\0&-1\end{psmallmatrix}.
\]
A \emph{Pauli string} ($\hat{P}$) on $n$ qubits is a tensor product of $n$ Pauli
matrices. For an $n$-site system, a Pauli string takes the form $\hat{P}_j = \pau_n \otimes \cdots \otimes \pau_1$, where each $\pau_k$ acts on a qubit Hilbert space $\Hilb_2$.
A Hamiltonian program can be rewritten as a weighted sum of Pauli strings: $\hat{H} = \sum_j r_j \hat{P}_j$, with $r_j\in \mathbb{R}^2$.

We write $Z(j)$ for $Z$ acting on qubit $j$ with identities
elsewhere, so $Z(0)\circ Z(1)$ is the Pauli string $Z \otimes Z$.  The
$4^n$ Pauli strings on $n$ qubits form a basis for all
$2^n \times 2^n$ matrices, so every Hamiltonian on $n$ qubits is a
weighted sum of Pauli strings.

\ignore{
The Jordan-Wigner transformation~\cite{Jordan1928} rewrites fermionic
operators as Pauli expressions:
\[
a(j) \;\mapsto\; Z(0)\circ\cdots\circ Z(j-1)\circ \tfrac12\bigl(X(j)+iY(j)\bigr),
\qquad
n(j) \;\mapsto\; \tfrac12\bigl(I - Z(j)\bigr).
\]

The $Z$ factors reproduce the fermionic sign.  This translation is the
trusted front end of \qsnd (\Cref{sec:intros}).  Applied to
\Cref{eq:hubbard-fermion} with $t=1$ and $U=2$, it produces the input
that \qsnd receives:
\[\small
\hat{H} =
 -\tfrac14\,(X(0)-iY(0))\circ Z(0)\circ(X(1)+iY(1))
 -\tfrac14\,Z(0)\circ(X(1)-iY(1))\circ(X(0)+iY(0))
 + 2\,\bigl(-\tfrac12 Z(0)\bigr)\circ\bigl(-\tfrac12 Z(1)\bigr).
\tag{2.2}\label{eq:hubbard-input}
\]
}

\begin{figure}[t]
\includegraphics[width=\textwidth]{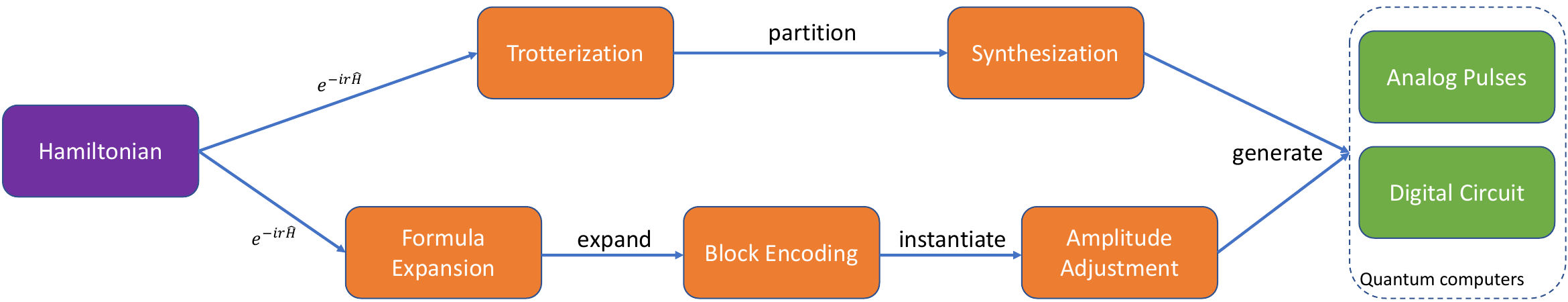}
\caption{Hamiltonian simulation compilation flow. The top flow is Trotter-based, and the bottom flow is BE-based.}
\label{fig:compilationprocess}
\end{figure}

\subsection{Trotter-Based Hamiltonian Simulation Compilation and QBlue}

Trotter-based Hamiltonian simulation compilation is based on the Trotter-Suzuki decomposition.
As shown in the top flow of \Cref{fig:compilationprocess}, the Trotter-based compiler has two steps: Trotterization and circuit synthesis.

To compile $\exp(\hat{H}_1+\hat{H}_2)$, we turn the exponential sum into a product of the two exponentials $\exp(\hat{H}_1)\circ\exp(\hat{H}_2)$.
However, the exponential of a sum equals the product of the exponentials only
when the terms commute.  Trotterization approximates
$\exp(-ir\hat{H})$ by a sequence of exponentials of single Pauli
strings.  The first-order product formula~\cite{Lloyd96,529425}
divides the time into $m$ steps,
\[
\exp\bigl(-ir(A+B)\bigr) \;\approx\;
\Bigl(\exp\bigl(-i\tfrac{r}{m}A\bigr)\circ\exp\bigl(-i\tfrac{r}{m}B\bigr)\Bigr)^{m},
\]
with an error of at most $\tfrac{r^2}{2m}\lVert [A,B] \rVert$ in
operator norm~\cite{PhysRevX.11.011020}, so the error shrinks as $m$
grows and vanishes when $A$ and $B$ commute.  Second-order
Trotterization~\cite{529425} applies the steps symmetrically, reducing
the error to $O(\frac{r^3}{m^2})$.  QDrift~\cite{Campbell_2019} samples terms
at random in proportion to their weights, so its cost depends on the
total weight of the Hamiltonian and not on its number of terms, and
MarQSim~\cite{cao2024marqsimreconcilingdeterminismrandomness} combines
deterministic and randomized orderings.

For the example with $r = \frac{\pi}{4}$, one step of first-order
Trotterization gives
\[
\exp\bigl(-\tfrac{i\pi}{4}\hat{H}\bigr) \;\approx\;
\exp\bigl(\tfrac{i\pi}{8}\,X(0)\circ X(1)\bigr)\circ
\exp\bigl(\tfrac{i\pi}{8}\,Y(0)\circ Y(1)\bigr)\circ
\exp\bigl(-\tfrac{i\pi}{8}\,Z(0)\circ Z(1)\bigr).
\tag{2.4}\label{eq:ex_final}
\]
The three Pauli strings in \Cref{eq:ex} commute pairwise, since each pair disagrees on both qubits, so in this example the approximation is an equality.  In general, the terms do not commute.  \qsnd then computes, for each of the four algorithms, an upper bound on the error as a function of its parameters, sets the parameters (the number of steps $m$, or the number of samples for the randomized algorithms) so that the bound stays below the user's $\epsilon$, and selects the algorithm with the smallest estimated cost (Appendix \ref{sec:compilationa}). For this example, it selects first-order Trotterization.

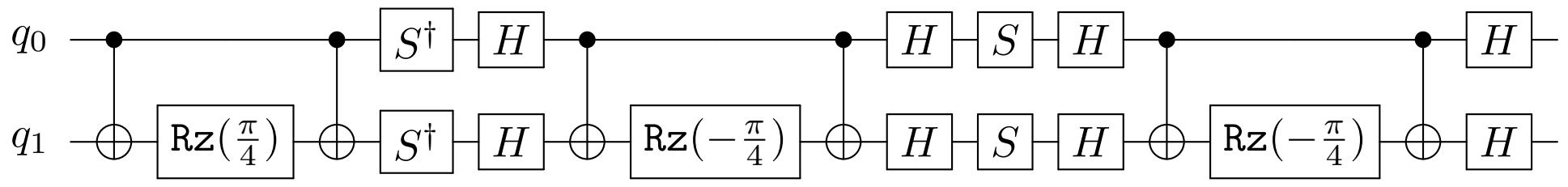
\begin{figure}[t]
\centering
\[
\Qcircuit @C=0.6em @R=0.8em {
  \lstick{q_0} & \ctrl{1} & \qw & \ctrl{1}
     & \gate{S^\dagger} & \gate{H} & \ctrl{1} & \qw & \ctrl{1} & \gate{H} & \gate{S}
     & \gate{H} & \ctrl{1} & \qw & \ctrl{1} & \gate{H} & \qw \\
  \lstick{q_1} & \targ & \gate{\cn{Rz}(\frac{\pi}{4})} & \targ
     & \gate{S^\dagger} & \gate{H} & \targ & \gate{\cn{Rz}(-\frac{\pi}{4})} & \targ & \gate{H} & \gate{S}
     & \gate{H} & \targ & \gate{\cn{Rz}(-\frac{\pi}{4})} & \targ & \gate{H} & \qw
}
\]
\caption{Digital circuit for $\exp(-i\frac{\pi}{4}\hat{H})$ in the
two-site model.  From left to right, the three blocks implement the
rotations about $Z(0)\circ Z(1)$, $Y(0)\circ Y(1)$, and
$X(0)\circ X(1)$ in \Cref{eq:ex_final}.}
\label{fig:hubbard-trotter-circuit}
\end{figure}

After Trotterization, \qsnd synthesizes the circuit to generate digital circuits or analog pulses.
Each factor of \Cref{eq:ex_final} is a rotation $\exp(-i\theta \hat{P})$
about the Pauli string $\hat{P}$.  On the digital backend, a rotation about
$Z(0)\circ Z(1)$ is a standard pattern,
\[
\exp\bigl(-i\theta\, Z(0)\circ Z(1)\bigr)
 = \cn{CNOT}_{0,1}\circ \bigl(I \otimes \cn{Rz}(2\theta)\bigr)\circ \cn{CNOT}_{0,1},
\qquad \cn{Rz}(\varphi) = \exp(-i\tfrac{\varphi}{2} Z),
\]
and rotations about other Pauli strings reduce to it by a change of
basis on each qubit: $H$ maps $X$ to $Z$, and $S^\dagger$ followed by
$H$ maps $Y$ to $Z$.  \Cref{fig:hubbard-trotter-circuit} shows the
resulting circuit, which implements \Cref{eq:ex_final} exactly.  The compilation to analog quantum computers realizes each rotation as a timed application of a
native interaction of the hardware instead of a gate sequence; the trapped-ion backend uses global
drives that couple several ions at once~\cite{Kyprianidis_2024,Richerme_2025}.

\subsection{BE-Based Algorithms}
\label{sec:trotter-overview}

In this paper, we focus mainly on \qsndl, which is based on BE-based algorithms, which is another way of compiling $\exp(-i \uapp{r}{\hat{H}})$ to a quantum circuit, where the subterms in $\hat{H}$ are unitary and can be directly implemented as quantum circuits, i.e., in a Pauli string form $\hat{H} = \sum_j r_j \hat{P}_j$, where $\hat{P}_j$ is unitary. Using the block encoding technique, we can turn the linear sum of Pauli strings into a sequence of control operations, each of which can be implemented directly as a quantum circuit.

The simplest BE-based algorithms is Taylor expansion, where we expand $\exp(-i r \hat{H})$ via the truncated Taylor series, $\sum_j \frac{- i r \hat{H}^k}{k!}$, which yields a summation operation, $\sum_k \theta_k \hat{P}_k$ for some Pauli-strings $\hat{P_k}$ and $\sum_k \slen{\theta_k}=1$, after normalization.
Such an operation can be implemented as a quantum circuit via the linear combination of unitaries (LCU) technique, implemented as a circuit $\cn{PRE};\cn{SELECT}; \cn{PRE}^{\dag}$, where the \cn{SELECT} circuit is demonstrated in \Cref{fig:hammingcircuit}.
\cn{PRE} prepares a superposition state and $\cn{SELECT}$ is a multiplexer circuit containing a sequence of controlled operations, each of which implements one Pauli string $\hat{P}_k$ as a quantum circuit, i.e., a controlled operation can be thought of as a quantum conditional, $\qif{x=k}{\hat{P}_k}$.
Details are given in \Cref{sec:compilation}.

\subsection{The Guarantee}

For a Hermitian Pauli-string Hamiltonian $\hat{H}$, a time $r$, and an
error bound $\epsilon$, \Cref{thm:compile-good} states that the compiled
program $U$ satisfies
$\lVert \llbracket U \rrbracket(\psi) - \llbracket \exp(-ir\hat{H})
\rrbracket(\psi) \rVert < \epsilon$ for every normalized state
$\psi$.  The proof ensures compiler correctness via the four BE-based algorithms, each with its own pipeline.
\qsndl bounds when it chooses the algorithm and its parameters.

\section{The QBlue Compilation Procedure}\label{sec:compilation}

This section describes the \qsndl compilation steps, shown in the bottom path of \Cref{fig:compilationprocess}, and explains how \rocq structures and verifies correctness proofs.

\subsection{The General Procedure}\label{sec:gen-compile}

The compiler transforms a Hamiltonian simulation expression $\eexp{\uapp{r}{\hat{H}}}$, where $\hat{H}$ is a \qsndl Hamiltonian program and $r$ is a time parameter, into a quantum circuit, approximating $\eexp{\uapp{r}{\hat{H}}}$ with a user-defined error bound $\epsilon$, executable on quantum hardware. 
We model compilation as a judgment of the form $(\hat{H}, r) \gg U$. The \qsndl compiler supports multiple compilation paths, helping users choose the best one from four BE-based algorithms.

The four different compilation algorithms have the same general pipeline, shown in the bottom path of \Cref{fig:compilationprocess}.
The pipeline contains the three compilation steps, listed below.

\begin{enumerate}
\item \textbf{Expansion Formula}: the primary step expands the Hamiltonian simulation operator $\eexp{\uapp{r}{\hat{H}}}$ to a linear sum of Pauli-strings.
These algorithms use two matrix expansion formulas: Taylor expansion and Jacobi-Anger expansion.

\item \textbf{Block Encoding}: A key component of the BE-based algorithm is block encoding, which implements a non-unitary matrix via a unitary operator executable on a quantum computer.
There are two main block encoding methods: LCU and quantum walk.
The given non-unitary operation can be rewritten as a linear sum operation, and the methods implement the summation via a multiplexer.
We discuss them in the next section.

\item \textbf{Amplitude Adjustment}: The amplitude adjustment step improves amplitude probabilities for correct Hamiltonian simulation, including oblivious amplitude amplification (OAA) and quantum signal processing.
The amplitude adjustment operation is implemented as a generic framework that takes the Hamiltonian simulation block encoding as input and produces a repeated pattern to improve the amplitudes.
\end{enumerate}

We now show the correctness theorem for compiling a Hamiltonian simulation of a constraint expression $\hat{H}$ to quantum circuits.

\begin{theorem}[Compilation Correctness]\label{thm:compile-good}\rm 
Given $\hat{H}$, a time period $r$, and a desired error $\epsilon$, we compile its simulation $\eexp{\uapp{r}{\hat{H}}}$ to a quantum circuit, as $(\hat{H},r)  \gg U$. For every state $\psi$, let $\psi_1=\denote{U}(\psi)$; the distance between the circuit and the simulation is bounded by $\epsilon$, as $\dabs{\psi_1 -\denote{\eexp{\uapp{r}{\hat{H}}}}(\psi)}<\epsilon$. 
\end{theorem}

\begin{proof}
    Fully mechanized proofs are done in Rocq for the four different versions of the four algorithms.
\end{proof}

\ignore{
\begin{table}[h]
{\footnotesize
\begin{tabular}{|l|c|}
\hline
Compiler Step & Theorems Being Verified \\

\specialrule{.2em}{.1em}{.1em} 
End-to-End Step & \Cref{thm:compile-good} for compiler correctness\\
\specialrule{.1em}{.05em}{.05em} 
\multirow{2}{*}{Canonicalization} & \Cref{thm:realness} for only real amplitudes in $\hat{H}$ \\
\cline{2-2}
& \Cref{thm:canonical-good} for canonicalizing any $\hat{H}$ to \Cref{def:canon-form}  \\
\specialrule{.1em}{.05em}{.05em} 
\multirow{4}{*}{Trotterization} & \Cref{thm:trotter-good} for Standard Trotterization \\
\cline{2-2}
  & \Cref{thm:trotter-good} for Second Order Trotterization \\
\cline{2-2}
    & \Cref{thm:trotter-good} for QDrift \\
\cline{2-2}
      & \Cref{thm:trotter-good} for MarQSim \\
\specialrule{.1em}{.05em}{.05em} 
\multirow{2}{*}{Synthesization} & \Cref{thm:decompose-good} for \Cref{fig:digitaldeco} \\
& \Cref{thm:analog-decompose-good} for \Cref{fig:analogdeco} \\
& \Cref{thm:analog-decompose-good} for \Cref{fig:inco} \\
\hline
\end{tabular}
}
\caption{Theorems being verified.}
\label{tab:proof-table}
\end{table}
}

The theorem states that the distance between the compiled circuit $U$ and the semantic definition $\eexp{\uapp{r}{\hat{H}}}$ is less than the user-defined bound $\epsilon$, and we relate this to other factors in the following section. 
Our verified compiler is end-to-end, with all theorems fully mechanized in \rocq.

\subsection{Individual BE-Based Algorithms}\label{sec:compile-ham-lcu}

This section discusses the four BE-based compilation algorithms based on the general framework above.

\subsubsection{Taylor Expansion based on LCU}

As mentioned in \Cref{sec:background}, one simple way to perform Hamiltonian simulation is to use a Taylor expansion. We discuss the algorithm based on the three steps in the general setting above.
The \textbf{expanding formula} step is explained in \Cref{sec:trotter-overview}, which expands $\eexp{\uapp{r}{\hat{H}}}$ as a truncated Taylor series. Below are the two other steps.

\paragraph{Block Encoding}\label{sec:taylor-block}
In the block encoding step, the Taylor expansion uses the LCU algorithm to implement a non-unitary operation on a quantum computer by composing many small unitaries.
Given a $2^n$ dimensional Hamiltonian $\hat{H}=\sum_j^m r_j \hat{P}_j$, applying to an $n$ qubit array $x$, we allocate a $\lift{\log{m}}$ ancilla qubit array $y$ and the LCU algorithm is a sequence of three components applying to $x$ and $y$, as $\cn{PRE};\cn{SELECT}; \cn{PRE}^{\dag}$, where $\cn{PRE}$ prepares the superposition state $\sum_j \sqrt{\frac{\slen{r_j}}{\lambda}}\,\ket{j}$ with $\lambda=\sum_j|r_j|$, and \cn{SELECT} applies $\hat{P}_j$ to $x$ when $y=j$ (with a sign flip when $r_j<0$).
We discuss the two components below.

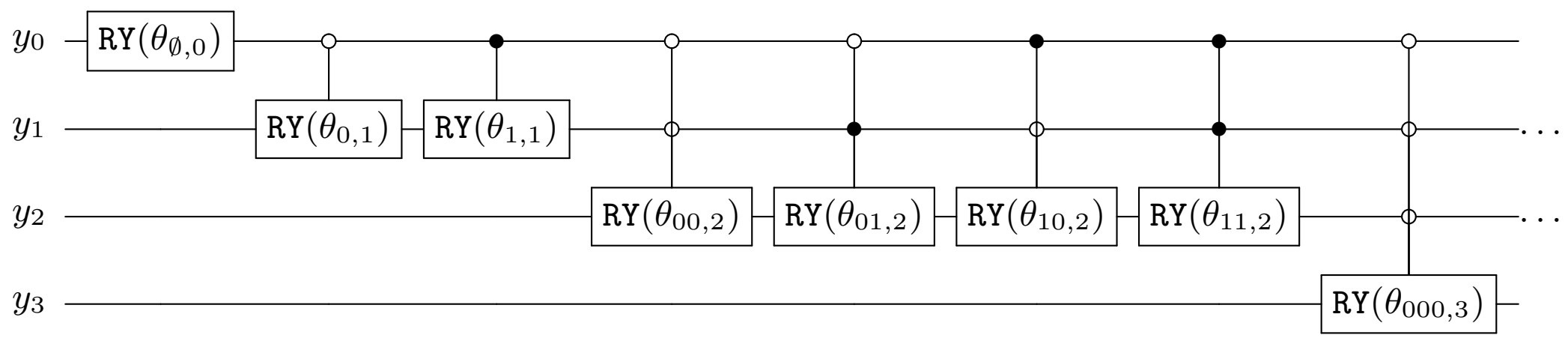
\begin{figure}[h]
\centering
\[\small
\Qcircuit @C=0.6em @R=0.8em {
  \lstick{y_0} & \gate{\cn{RY}(\theta_{\emptyset,0})} & \ctrlo{1} & \ctrl{1}        & \ctrlo{2} & \ctrlo{2} & \ctrl{2} & \ctrl{2} & \ctrlo{3} & \qw & \\
  \lstick{y_1} & \qw & \gate{\cn{RY}(\theta_{0,1})} & \gate{\cn{RY}(\theta_{1,1})}  & \ctrlo{1} & \ctrl{1} & \ctrlo{1} & \ctrl{1} & \ctrlo{2} & \qw & \dots \\
   \lstick{y_2} & \qw & \qw & \qw  & \gate{\cn{RY}(\theta_{00,2})} & \gate{\cn{RY}(\theta_{01,2})} & \gate{\cn{RY}(\theta_{10,2})} &\gate{\cn{RY}(\theta_{11,2})} & \ctrlo{1} &  \qw & \dots \\ 
     \lstick{y_3} & \qw & \qw & \qw & \qw & \qw & \qw & \qw & \gate{\cn{RY}(\theta_{000,3})}  &  \qw & \\
}
\]
\caption{The \cn{PRE} circuit based on binary tree decomposition and \cn{RY} gates.}
\label{fig:tlcu-circuit}
\end{figure}

\begin{figure}[t]
{
$\begin{array}{c}
  \Qcircuit @C=0.5em @R=0.5em {
    &                     &  & \ctrlo{6} & \ctrlo{6} & \qw & \qw & \qw & \ctrl{6} & \qw &  \\
    &                     &  & \ctrlo{5} & \ctrlo{5} & \qw & \qw & \qw & \ctrl{5} & \qw &   \\
    & \push{y:\ket{0}_{\lift{\log{m}}}\quad} &   & & & &  & & & & \\
    &                     &  & &  & & \dots & & & & \\
    &                     & & & & & & & & &  \\
    &                     & & \ctrlo{1} & \ctrl{1} & \qw & \qw & \qw & \ctrl{1} & \qw & \\
    &                     & & \multigate{4}{\hat{P}_0} & \multigate{4}{\hat{P}_1} & \qw & \qw & \qw & \multigate{4}{\hat{P}_m} & \qw &  \\
    & \push{x:\ket{0}_n\quad } & & \ghost{\hat{P}_0} & \ghost{\hat{P}_1} & \qw & \qw & \qw & \ghost{\hat{P}_m} & \qw &   \\
    &                     &   &                                 &                                     &     & \dots &     &                                         & & \\
    &                     &        &                                     &                                     &     &       &     &                                         & &  \\
    &                     &   & \ghost{\hat{P}_0} & \ghost{\hat{P}_1} & \qw & \qw   & \qw & \ghost{\hat{P}_m} & \qw 
    \gategroup{1}{3}{6}{3}{1em}{\{}
    \gategroup{7}{3}{11}{3}{1em}{\{}
    }
    \end{array}
$
}
\caption{The \cn{SELECT} circuit based on multiplexer.}
\label{fig:hammingcircuit}
\end{figure}
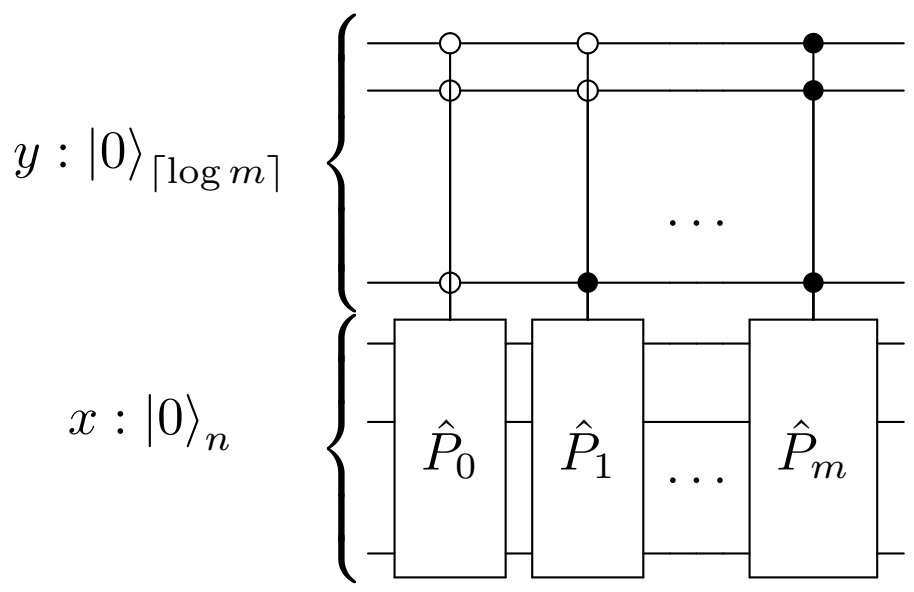

\Cref{fig:tlcu-circuit} shows the \cn{PRE} circuit that prepares a superposition state for $\lift{\log{m}}$ ancilla qubits via the Grover-Rudolph algorithm, which turns a state-preparation problem into a binary-tree decomposition problem.
In the circuit generation, for a $\lift{\log{m}}$ qubit state, the $k$-th qubit ($k\in[0,\lift{\log{m}})$) has $2^k$ multiplexers, with the controlling gate being an \cn{RY} gate, where the $k$-th qubit has multiplexers with $k$ nodes, so the first qubit has a pure \cn{RY} gate ($k=0$).
To better describe the circuit, we assume $k$-length binary bitstrings $b \in \{0,1\}^k$, with $k\in[0,\lift{\log{m}})$, i.e., the $k$-th qubit requires $2^k$ bitstrings, each with $k$ bits.
We record the angle value $\theta_{b,k}$ in the $\cn{RY}(\theta_{b,k})$ gate for the $k$-th qubit as follows (assuming $t=\lift{\log{m}}$).

{\small
\[
\theta_{b,k} \triangleq 2 \arctan2\Big(\sqrt{\textstyle\sum_{j=b\cdot 2^{t-k}+2^{t-k-1}}^{(b+1)2^{t-k}-1} |r_j|},\ \sqrt{\textstyle\sum_{j=b\cdot 2^{t-k}}^{b\cdot 2^{t-k}+2^{t-k-1}-1} |r_j|}\Big)
\]
}

As an instance, $\theta_{\emptyset,0}$ is equal to $2 \arctan2\big(\sqrt{\sum_{j=2^{t-1}}^{2^{t}-1} |r_j|},\sqrt{\sum_{j=0}^{2^{t-1}-1} |r_j|}\big)$.
Given the initial state $\ket{0}_{\lift{\log{m}}}$, the whole circuit prepares $\sum_j \sqrt{\frac{\slen{r_j}}{\lambda}}\,\ket{j}$; each uniformly controlled \cn{RY} is compiled into \cn{RY} and \cn{CNOT} gates.

As shown in \Cref{fig:hammingcircuit}, the \cn{SELECT} operation is a multiplexer that encodes the application of the $j$'s Pauli string as a controlled operation, with the controlled node on $j$.
The encoding writes the index $j$ as a $\lift{\log{m}}$-bit string and processes each bit: $1$ is encoded as a standard controlled node (black node), and $0$ as a \cn{X} gate with a controlled node (white node).
\Cref{fig:hammingcircuit} shows an example of controlled-node encoding, where the white and black nodes represent conditions on the $0$ and $1$ bits, respectively.

\myparagraph{Amplitude Adjustment}\label{sec:oaa-compile}
In the amplitude adjustment step, Taylor expansion uses the OAA algorithm to amplify the amplitude of a target state.
Given the \cn{LCU} circuit above applied to the two registers $x$ and $y$, we define a reflection operation to reflect on an all-$0$ bitstring, listed below.

{\small
\[
R \triangleq \cn{X}(u); \cn{H}(u); \qif{y=0}{\cn{X}(u)};\cn{H}(u);\cn{X}(u)
\]
}

Here, $\qif{y=0}{\cn{X}(u)}$ is a quantum conditional, implemented as a multiplexer querying whether the quantum array $y$ is all $0$ ($y=0$).

A one-step OAA algorithm is implemented as, $\cn{OAA} \triangleq \cn{LCU};R;\cn{LCU}^{\dag};R;\cn{LCU}$, where $R$ is defined above and \cn{LCU} is defined in Section \ref{sec:taylor-block}.
In general, boosting an amplitude $s$ takes $O(\frac{1}{s})$ rounds; we pad each segment's amplitude to exactly $\frac12$ with one extra \cn{RY} rotation, so a single round gives the target exactly.
The OAA framework is a top-level framework that wraps the LCU operation developed above, as $\cn{OAA}(\cn{LCU})$.

\myparagraph{Efficiency}
Eventually, we verify \Cref{thm:compile-good} for the Taylor expansion algorithm, establishing the following relations among error bounds, time periods, and steps.
For $M \geq 1$ simulation segments, the mechanized error-budget lemma establishes that a local error bound $\delta \leq \frac{\epsilon}{M}$ implies $M\delta \leq \epsilon$.
This provides the error budget required to compose the Taylor simulation segments.

\subsubsection{Qubitization with Quantum Walk}
Qubitization builds a walk operator from the block encoding whose powers apply Chebyshev polynomials of $\frac{\hat{H}}{\lambda}$; combining those powers with the right weights gives $\eexp{\uapp{r}{\hat{H}}}$.
The algorithm follows the pipeline in \Cref{fig:compilationprocess}.

\myparagraph{Expansion Formula}
The qubitization with quantum walk method approximates the time evolution with the Jacobi--Anger expansion for $\hat{H}=\sum_j r_j \hat{P}_j$, as listed below.
\begin{equation}
  \eexp{\uapp{r}{\hat{H}}} \;\approx\; J_0(-\lambda r) + 2\sum_{k=1}^{K} i^k J_k(-\lambda r)\, T_k(\frac{\hat{H}}{\lambda})
  \label{eq:jacobi}
\end{equation}

Here, $\lambda=\sum_j|r_j|$. We truncate the expansion at $K$, and $J_k(r)$ is the Bessel function, defined as follows.

{
\[
J_k(r) \triangleq \frac{1}{\pi} \int_0^{\pi}\cos(k\theta - r \sin(\theta)) d\theta
\]
}

Note that the function arguments $k$ and $r$ are numbers; thus, the computation cost is minimal.
The $T_k(\hat{H})$ term approximates $\cos(\hat{H})$ using truncated Chebyshev polynomials of the first kind, whose recursive formula is listed below.

{
\[
T_0(\hat{H}) \triangleq I
\qquad
T_1(\hat{H}) \triangleq \hat{H}
\qquad
T_{n+1}(\hat{H}) \triangleq 2\hat{H}\circ T_n(\hat{H}) - T_{n-1}(\hat{H})
\]
}

In \Cref{eq:jacobi}, we need to adjust $\hat{H}$'s amplitude by performing $\frac{\hat{H}}{\lambda}$ because Chebyshev polynomials are bounded only on $[-1,1]$.
The expansion converges quickly. Each $T_k$ is bounded by $1$, so cutting the sum at $K$ costs at most $\sum_{k>K}2|J_k(\tau)|$, and $J_k(\tau)$ falls off faster than exponentially once $k$ exceeds $\tau$. In practice, we split the time $r$ into $M=\lceil\lambda r\rceil$ segments, so each segment has $\tau\le 1$, and pick the smallest $K$ whose tail is below $\frac{\epsilon}{4M}$. That usually leaves only a handful of terms per segment.

\myparagraph{Block Encoding}
We build the walk from the same \cn{PRE} and \cn{SELECT} as in Section~\ref{sec:taylor-block}, applied to the $x$ and $y$ registers:

{
\[
\cn{WALK} \triangleq \cn{SELECT};\cn{PRE}^{\dag};R;\cn{PRE}
\]
}

Here $R$ is the reflection about $\ket{0}$ on $y$ defined above, so $\cn{PRE}^{\dag};R;\cn{PRE}$ is the reflection about $\ket{G}=\cn{PRE}\ket{0}$, an amplitude amplification diffusion operation~\cite{Brassard_2002}.
Its powers give the Chebyshev terms directly: $\bra{G}\cn{WALK}^k\ket{G}=(-1)^k T_k(\frac{\hat{H}}{\lambda})$, so the $k$-th term of \Cref{eq:jacobi} is one power of \cn{WALK}.
Here, $\cn{WALK}^k$ means repeating $\cn{WALK}$ $k$ times.

\myparagraph{Amplitude Adjustment}
We combine $\cn{WALK}^0,\dots,\cn{WALK}^K$ with one more LCU layer. An outer register is prepared with amplitudes $\sqrt{\frac{\slen{\beta_k}}{B}}$, where $\beta_0=J_0(\tau)$, $\beta_k=2J_k(\tau)$, and $B=\sum_k|\beta_k|$; the phases $i^k$ and the signs of $\beta_k$ are applied on that register, and its bits control $\cn{WALK}^{2^j}$.
A padding rotation sets the success amplitude to exactly $\frac12$, so one OAA round, applied to this whole segment rather than to a single \cn{WALK}, makes the segment deterministic.

\myparagraph{Efficiency}
We verify \Cref{thm:compile-good} for the qubitization with quantum walk algorithm, establishing the following relations among error bounds, time periods, and steps.
For a simulation time $r$ divided into $M$ segments, each segment simulates an evolution of duration $\frac{r}{M}$.
Under the segment-level correctness and non-expansiveness assumptions, we prove in Rocq that the accumulated
simulation error is bounded by $M\delta$, where $\delta$ is the error bound of a single compiled segment.
Consequently, choosing $\delta \leq \frac{\epsilon}{M}$ guarantees an overall error of at most $\epsilon$.

\subsubsection{Qubitization with Quantum Signal Processing}
QSP builds on qubitization: instead of combining powers of \cn{WALK} by LCU, it applies a polynomial of \cn{WALK} directly, so no OAA is needed.
We use generalized QSP~\cite{motlagh2024}, which interleaves controlled walk steps with rotations on one extra qubit $u$:

{\small
\[
\cn{QSP}(\Phi) \triangleq \cn{PRE};\,R_u(\phi_0);\,\big(\cn{C}_u\cn{WALK};\,R_u(\phi_1)\big);\dots;\big(\cn{C}_u\cn{WALK};\,R_u(\phi_{2K})\big);\,(\cn{WALK}^{\dag})^{K};\,\cn{PRE}^{\dag}
\]
}

Here $\cn{C}_u\cn{WALK}$ applies \cn{WALK} when $u=0$, and each $R_u(\phi_j)$ is a single-qubit rotation on $u$.
On each eigenvector of $\hat{H}$, \cn{WALK} acts as a phase $z$, so the $2K$ controlled steps apply a degree-$2K$ polynomial $P(z)$, and the $K$ inverse walks shift it to $z^{-K}P(z)$, which matches the Jacobi--Anger sum from $-K$ to $K$.
The angles come from the coefficients of $P$ by peeling off one rotation at a time: with a complementary polynomial $Q$ satisfying $|P|^2+|Q|^2=1$ on the unit circle, each layer is chosen so that undoing it lowers the degree of $(P,Q)$ by one~\cite{motlagh2024}.

\begin{algorithm}[h]
\caption{QSP Phase-Factor Optimization Solver (\textsf{QSPBFGS})}
\label{alg:qsp_solver}
\begin{algorithmic}[1]
\REQUIRE Target real polynomial $f$ of degree $d$ with $|f|<1$ on $[-1,1]$, tolerance $\epsilon$.
\ENSURE Symmetric phase vector $\Phi=(\phi_0,\dots,\phi_d)$ with $\phi_k=\phi_{d-k}$.
\STATE Let $\tilde{d} = \lceil\frac{d+1}{2}\rceil$ and choose the nodes $x_j = \cos\left(\frac{(2j-1)\pi}{4\tilde{d}}\right)$, $j=1,\dots,\tilde{d}$ (the positive roots of $T_{2\tilde{d}}$).
\STATE Define $L(\hat{\Phi}) = \frac{1}{\tilde{d}}\sum_{j}\big|\mathrm{Re}\,\langle 0|U_{\Phi}(x_j)|0\rangle - f(x_j)\big|^2$, where $\hat{\Phi}$ holds the $\tilde{d}$ free phases.
\STATE Set $\hat{\Phi}^{(0)} = (\frac{\pi}{4},0,\dots,0)$, initialize the inverse Hessian approximation $B_0$, and set $t = 0$.
\WHILE{$L(\hat{\Phi}^{(t)}) > \epsilon$ and L-BFGS still makes progress}
    \STATE Compute $\nabla L(\hat{\Phi}^{(t)})$ and update $\hat{\Phi}^{(t+1)}$ with an L-BFGS step; $t \leftarrow t + 1$.
\ENDWHILE
\STATE Refine $\hat{\Phi}^{(t)}$ with Newton steps on $\mathrm{Re}\,\langle 0|U_{\Phi}(x_j)|0\rangle = f(x_j)$ until the residual is below $\epsilon$.
\STATE \textbf{return} the full symmetric $\Phi$ built from $\hat{\Phi}^{(t)}$.
\end{algorithmic}
\end{algorithm}

Finding the phase angles $\Phi$ is the hard part for high-degree polynomials, so we use the optimization-based solver of Dong et al.~\cite{donglin}, summarized in Algorithm~\ref{alg:qsp_solver}.
Instead of computing the phases one at a time from the polynomial's roots, which becomes numerically unstable at high degree, the solver treats them as unknowns and fits them directly: it evaluates the QSP sequence at the Chebyshev nodes in step~1 and minimizes the squared difference from the target polynomial with L-BFGS. Because the phases are symmetric, only half of them are free, and starting from the trivial phases $(\frac{\pi}{4},0,\dots,0)$, the optimization converges reliably even for degrees in the hundreds; in our implementation, a few Newton steps then polish the result to the required tolerance.
We use this solver for the QSVT phases (\Cref{sec:qsvt}); our QSP path computes its angles by layer stripping instead~\cite{motlagh2024}.

\myparagraph{Efficiency and Error Bounds}
To establish the relationship between the truncation degree, simulation time, and approximation error, let $r$ denote the total simulation time, $\lambda=\sum_j \slen{r_j}$ the LCU
normalization factor, and $\epsilon$ the target error bound. The QSP compilation divides the evolution into
\[
    M=\max\{1,\lceil \frac{\lambda r}{10}\rceil\}
\]
segments, assuming $r\geq 0$, each with normalized simulation time $\tau=\frac{\lambda r}{M}$. Let $d$ denote the Jacobi--Anger truncation order. The corresponding generalized QSP polynomial has degree $2d$.

\begin{enumerate}

    \item \textbf{Spectral Decomposition:}
    For the normalized Hamiltonian $x=\frac{\hat{H}}{\lambda}$, whose spectrum lies in  $[-1,1]$, the time-evolution function is expressed using the Jacobi--Anger expansion:
    \[
        \eexp{\uapp{\tau}{x}}
        =
        J_0(\tau)T_0(x)
        +
        2\sum_{k=1}^{\infty}
        (-i)^k J_k(\tau)T_k(x),
    \]
    where $T_k$ is the Chebyshev polynomial of the first kind and $J_k$ is the Bessel function of the first kind.

    \item \textbf{Series Truncation:}
    Truncating the expansion at degree $d$ introduces a polynomial approximation error bounded by
    \[
        \delta_{\mathrm{poly}}(d,\tau)
        =
        2\sum_{k=d+1}^{\infty}|J_k(\tau)|.
    \]
    Since $|T_k(x)|\leq 1$ for $x\in[-1,1]$, this Bessel tail provides a uniform bound on the truncation error:
    \[
        \sup_{x\in[-1,1]}
        |\eexp{\uapp{\tau}{x}}-P_d(x)|
        \leq \delta_{\mathrm{poly}}(d,\tau),
    \]
    where
    \[
        P_d(x)=J_0(\tau)T_0(x)
        +2\sum_{k=1}^{d}(-i)^kJ_k(\tau)T_k(x).
    \]

    \item \textbf{Truncation Degree and Circuit Steps:}
    The compiler selects the truncation order $d$ using the local tolerance
    \[
        \epsilon_0=\frac{\epsilon}{4M}.
    \]
    The intended truncation condition is
    \[
        \delta_{\mathrm{poly}}(d,\tau)
        \leq \epsilon_0.
    \]
    The generalized QSP construction forms a degree-$2d$ polynomial and implements
    $2d$ controlled quantum-walk operations and $d$ inverse-walk operations per segment. Thus, the circuit construction relates the number of walk operations directly to the selected truncation order.

\end{enumerate}

To connect these approximation bounds with the mechanized error analysis, let $\delta_{\mathrm{phase}}$ denote the error budget associated with phase or angle synthesis, including the complementary-polynomial construction, and let $\delta_{\mathrm{impl}}$ denote the remaining implementation error budget, including polynomial
scaling and circuit realization errors. All error budgets are defined per segment.

The mechanized Rocq lemma
\texttt{component\_budget} establishes that,
for $M\geq 1$ and nonnegative error budgets,
\[
    \delta_{\mathrm{poly}}(d,\tau)
    +\delta_{\mathrm{phase}}
    +\delta_{\mathrm{impl}}
    \leq \frac{\epsilon}{M}
\]
implies
\[
    M\left(
        \delta_{\mathrm{poly}}(d,\tau)
        +\delta_{\mathrm{phase}}
        +\delta_{\mathrm{impl}}
    \right)
    \leq \epsilon.
\]

The mechanized result therefore establishes the arithmetic relationship between the local error
budget, the number of simulation segments, and the global tolerance. When the individual components
bound the complete per-segment semantic error and the segment-composition assumptions are satisfied,
this yields an accumulated simulation error bounded by $\epsilon$. The corresponding local approximation and circuit-level correctness obligations must be established separately.

\subsubsection{Quantum Singular Value Transformation}\label{sec:qsvt}
QSVT also applies a polynomial of $\frac{\hat H}{\lambda}$, but it puts the phases on the block encoding itself rather than on a controlled walk.
The expansion formula uses the Jacobi-Anger expansion with Bessel coefficients.
The block encoding step uses the LCU block $U=\cn{PRE};\cn{SELECT};\cn{PRE}^{\dag}$ from Section~\ref{sec:taylor-block}.
The key difference happens in the amplitude adjustment step.

\myparagraph{Amplitude Adjustment}
QSVT alternates $U$ and $U^{\dag}$ with projector phases
$\eexp{{\phi_j}{(2\Pi-I)}}$, where $\Pi=\ket{0}\bra{0}_y$; each phase is a
multi-controlled flag on one ancilla followed by an \cn{RZ}. With the right
angles $\phi_1,\dots,\phi_d$, the top-left block of the result is a degree-$d$
polynomial of $\frac{\hat H}{\lambda}$. One sequence can only produce a polynomial of
fixed parity, so we build two: an even one for the cosine part and an odd
one for the sine part of the Jacobi--Anger sum. One extra qubit takes the
real part of each (running the sequence with $\phi$ and $-\phi$ and
averaging), and a second one combines the two, which gives
$\frac12(P_{\cos}+iP_{\sin})\approx\frac12 \exp(\frac{\uapp{-i\tau}{\hat{H}}}{\lambda})$.
Since the amplitude is already $\frac12$, one round of OAA makes the
segment deterministic. The angles come from Algorithm~\ref{alg:qsp_solver}.

\myparagraph{Efficiency}
With the above setting, we establish the following relationship between the steps and the error bounds.
For a simulation time $r$, the QSVT compilation constructs $M=\lceil\frac{\lambda r}{50}\rceil$ segments, each targeting an evolution
of duration $\frac{r}{M}$, with normalized segment duration $\tau = \frac{\lambda r}{M}$.
The compiler sets both the polynomial truncation tolerance and the polynomial
scaling parameter $\eta$ to $\frac{\epsilon}{8M}$. Under the segment-level correctness
and non-expansiveness assumptions, we prove in Rocq that the accumulated
simulation error is bounded by $M\delta$, where $\delta$ denotes the error of a
complete QSVT segment after OAA. Consequently, requiring $\delta \leq \frac{\epsilon}{M}$
guarantees an overall simulation error of at most $\epsilon$.
The local bound follows from the parameters. Truncation and the
$(1-\eta)$ scaling each cost at most $\frac{\epsilon}{8M}$, so the encoded
operator is within $\gamma=\frac{\epsilon}{4M}$ of $\exp(\frac{\uapp{-i\tau}{\hat{H}}}{\lambda})$, and
one OAA round keeps the error within $\gamma+2\gamma^2\le\frac{\epsilon}{2M}$,
which meets $\delta\le\frac{\epsilon}{M}$ with a factor of two to spare.
The phases are computed numerically and checked against their target
polynomials; over our benchmarks the largest residual is $8.5\times10^{-13}$.

 \section{Evaluation}
\label{sec:arith-oqasm}

This section presents an experimental evaluation of the \qsndl. We first define our benchmark suite of quantum programs and the state-of-the-art baseline frameworks for comparison. We then present our results, structured to answer the following research questions:

\begin{itemize}
\item RQ1: Applicability: Is \qsndl applicable to compile BE-based quantum simulation programs to quantum circuits, within the allowed error bounds?
\item RQ2: Effectiveness: Compared to other frameworks, does \qsndl produces effective quantum circuits for a wide range of programs?
\item RQ3: Correctness and Reliability: Compared to other frameworks, can \qsndl produces correct circuits with reliable results?
\end{itemize}

\subsection{Experimental Setup}

\begin{table}[h]
{\footnotesize
\begin{tabular}{|l|c|c|c|c|c|}
\hline
Names & \# & Ave. Qubits \# & Qubits Range & Ave. Pauli strings \# & Range Pauli strings \# \\
\hline
\texttt{Small Set} & 220 & 10 & 4 - 32 & 27 & 6 - 80 \\
\hline
\texttt{Medium Set} & 36 & 13 & 8 - 40 & 518 & 200 - 1884 \\
\hline
\texttt{Large Set} & 23 & 19 & 12 - 72 & 2440 & 2000 - 368021 \\
\hline
\end{tabular}
}
\caption{The experimental data information. The $\#$ column refers to the number of test cases.}
\label{tab:data-table}
\end{table}

\myparagraph{Benchmark Suite}
\label{sec:benchmark} 
To evaluate \qsndl, we create a benchmark of 279 Hamiltonian simulation programs from real-world applications simulating physical and chemical systems related to electron behavior. 
The dataset is collected from a quantum material simulation project \cite{Yao_2021,chen2025quantumclassicalembeddingghostgutzwiller}, Genesis \cite{10.1145/3695053.3731065}, and test programs from previous Hamiltonian simulation projects \cite{cao2024marqsimreconcilingdeterminismrandomness,10.1145/3503222.3507715}.
It includes physical models, such as 1D Transverse-Field Ising and Hubbard models, as well as quantum chemistry models, such as particle simulations of $BeH_2$, the $eL42$ protein, and ethylene. The full list is available in our artifact, and we summarize their properties in \Cref{tab:data-table}. We run all experiments on $276$ of them: the three largest programs (two benzene Hamiltonians with $368{,}021$ Pauli strings and a $C_2H_4$ fragment with $152{,}371$) exceed our per-job time limit of two hours, since computing their step counts alone takes hours to days.

When selecting benchmarks, we partition the dataset by Pauli string length into three sets, small, medium, and large, to keep the dataset diverse and evaluate \qsndl in a diverse environment. Programs in the small set are traditional benchmark programs for quantum computing-based Hamiltonian simulations, while programs in the large set, e.g., benzene and $C_2H_4$, serve as small components in real-world chemistry applications.
When testing the dataset, we perform simulation with time $r$ to be $\frac{\pi}{16}$, $\frac{\pi}{4}$, $\pi$, and $4\pi$, as well as error bounds set to be $0.0002$, $0.002$, $0.02$, and $0.1$.

\myparagraph{Baseline Systems} We compare \qsndl's efficiency against a baseline of a state-of-the-art framework, PennyLane \cite{bergholm2022pennylaneautomaticdifferentiationhybrid}, containing a quantum simulation framework with BE-based algorithms. 
PennyLane is an industrial compiler, similar to other works such as Qiskit Nature \cite{qiskit2024}, and it mainly focuses on quantum chemistry and quantum particle applications, such as computing ground energy states.
We want to compare it, because it is the representative BE-based quantum simulation compilers in industry, which provide the ability of compiling a real-world application to quantum computers, rather than focusing on some subroutines \cite{10.1145/3470496.3527394} or using other kinds of machines \cite{10.1145/3695053.3731065} and compilation methods \cite{arrazola2023differentiablequantumcomputationalchemistry,Andrew2018}.

\myparagraph{Hardware and Implementation}\label{sec:hardware} We extracted the \qsndl compiler from Rocq to OCaml and performed the evaluation on an Ubuntu computer with an 8-core 13th-gen Intel i9 processor and 16GB of DDR5 memory. 

\subsection{RQ1: Applicability}\label{sec:datacomparison}

\begin{figure}[h!]
    \centering
    \includegraphics[width=0.8\linewidth]{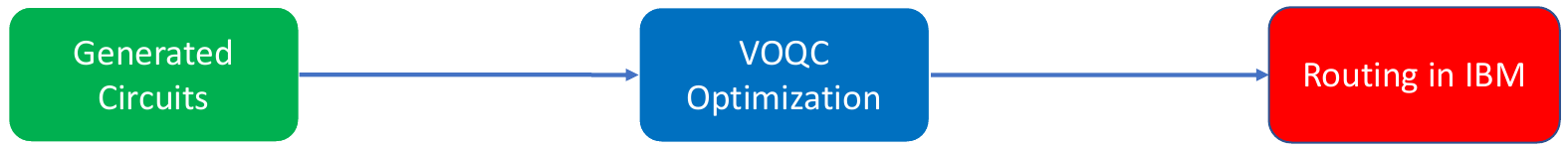}
    \caption{\qsndl compilation steps to the IBM quantum computer after the BE-algorithm step.}
    \label{fig:steps-after-syn}
\end{figure}

\begin{figure}[t]
    \includegraphics[width=1\linewidth]{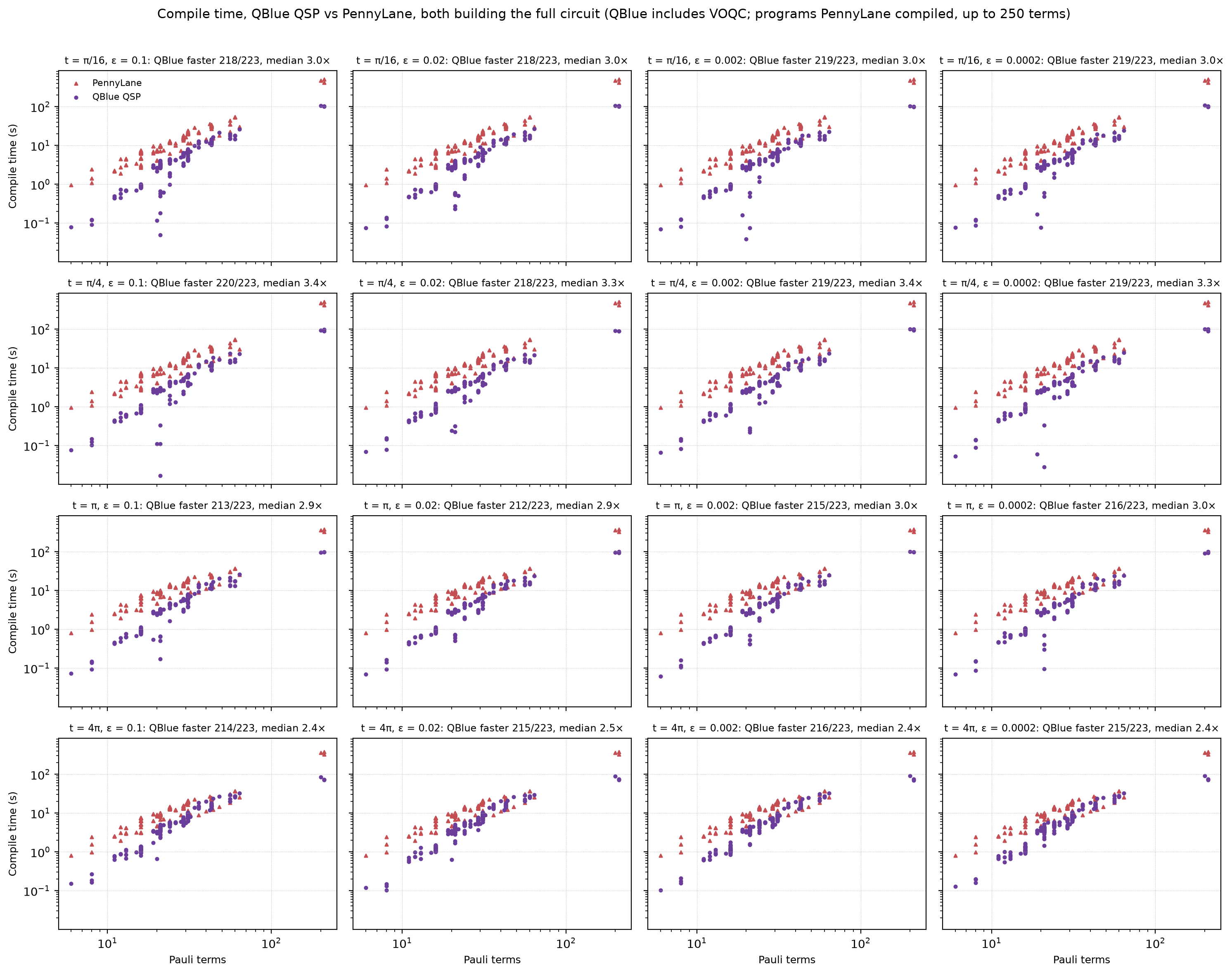}
    \caption{Compilation time comparison between QBlue and PennyLane; error bounds: $0.0002$, $0.002$, $0.02$, and $0.1$, and simulation time periods: $\pi/16$, $\pi/4$, $\pi$, $4\pi$.}
    \label{fig:compilation-time-comparison}
\end{figure}

To address the first research question, we evaluate \qsndl by compiling the simulations of the Hamiltonian programs on the three datasets for IBM digital circuits and comparing compilation differences between \qsndl and PennyLane.

Our compilation follows the compilation pipeline in \Cref{fig:compilationprocess}, with two additional steps following the pipeline: 1) we apply a standard quantum circuit optimizer, VOQC \cite{VOQC}, to optimize the generated circuits, and 2) we perform the circuit routing steps targeting the IBM quantum computer structure by inserting necessary gates, such as \cn{SWAP} gates, to permit the execution of the generated circuits on the IBM machine. \Cref{fig:compilation-time-comparison} compares the compilation time of \qsndl with PennyLane for input error bounds $0.0002$, $0.002$, $0.02$, and $0.1$, and simulation time periods $\frac{\pi}{16}$, $\frac{\pi}{4}$, $\pi$, and $4\pi$. \qsndl compiles faster on $212$--$220$ of the $223$ programs that PennyLane is able to compile in each setting, with a median speedup of $2.4$--$3.4\times$. Additionally, PennyLane failed to compile 53 programs; all failed programs have more than $250$ Pauli strings.

The results show that \qsndl can produce Hamiltonian simulation results for all programs across settings in a reasonable amount of time.
This shows that \qsndl is applicable across different programs.

\subsection{RQ2: Effectiveness}\label{sec:works-comparison}

\begin{figure}[t]
{\hspace*{-2em}
    \includegraphics[width=1.05\linewidth]{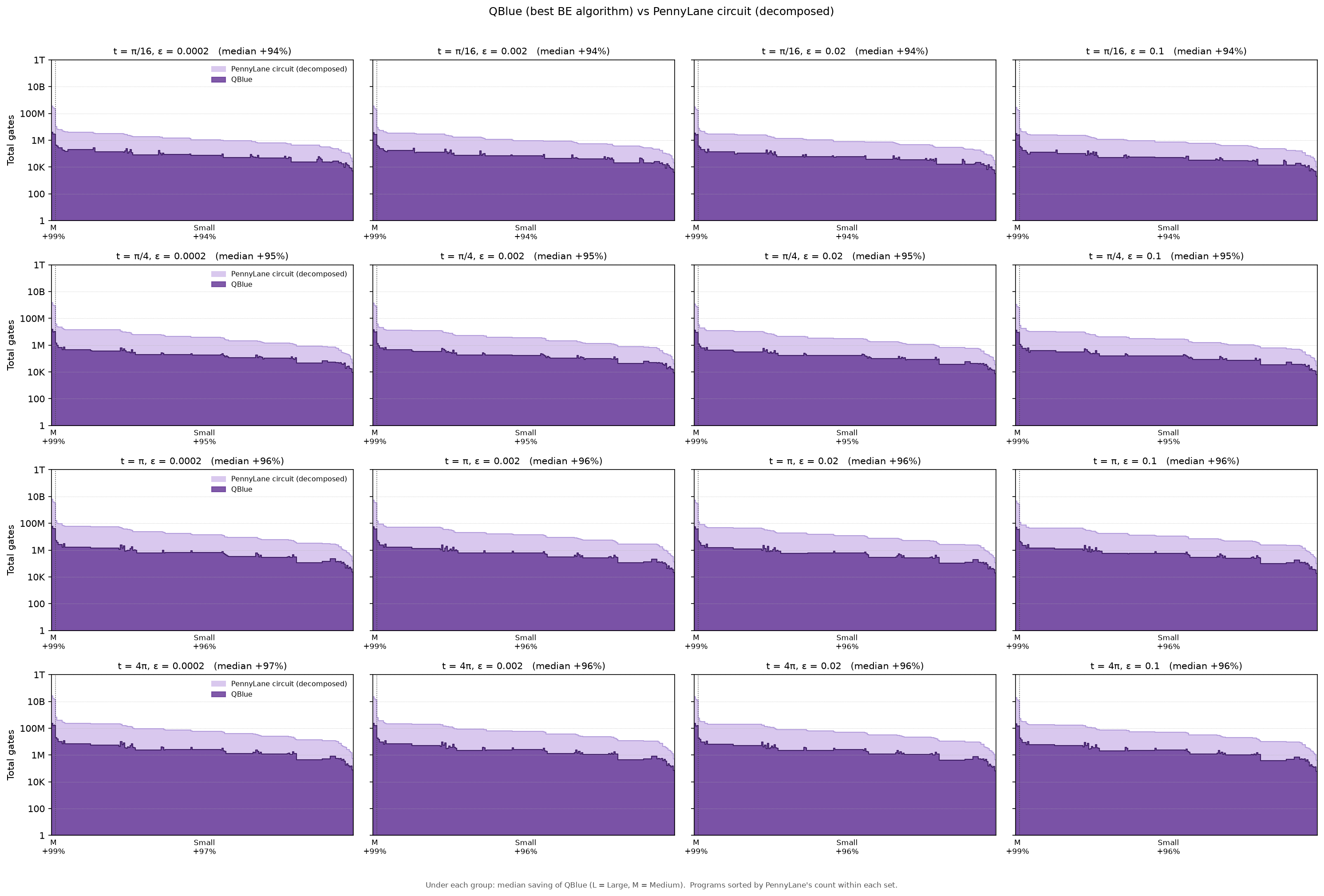}
    }
\caption{\qsndl optimization reduction compared to PennyLane for error bounds:  $0.0002$, $0.002$, $0.02$, and $0.1$, and simulation time: $4\pi$, $\pi$, $\frac{\pi}{4}$, and $\frac{\pi}{16}$. In each figure, the upper panel compares against PennyLane. We compare the best-performing algorithm for each program.}
\label{tab:splitting-step1}
\end{figure}

We now examine the effectiveness of \qsndl by comparing it with PennyLane, since compilation time is usually the least concern in Hamiltonian simulation; optimizing digital gates is the main factor.
Mainly, we compare \qsndl's overall compilation performance, in terms of circuit gate savings, against PennyLane. We only compare the programs that PennyLane is able to compile.
To compare overall performance, for each program we input the desired input error bounds and simulation time periods, and let the two platforms select the optimal result circuit.

We present the experiment results in \Cref{tab:splitting-step1} for the error bounds and time periods in Section \ref{sec:benchmark}.
The results show that \qsndl significantly outperforms PennyLane in compilation performance.
The compilation results show that \qsndl performs even better on medium programs than on small ones.
On average, \qsndl yields roughly $98.9\%$ and $95.6\%$ gate reduction on the medium and small sets, respectively, while we do not compare the large set because PennyLane cannot compile programs in that set.
\qsndl produces fewer gates than PennyLane in all $3{,}568$ cases we compared.
The largest gaps appear on medium-set programs at long time periods and tight error bounds. For the SYK model at $4\pi$ and $\epsilon=0.0002$, \qsndl produces a circuit about $\frac{1}{108}$ the size of PennyLane's ($1.6\times10^{8}$ vs.\ $1.7\times10^{10}$ gates), and for the B3 impurity model it is about $\frac{1}{107}$ ($2.4\times10^{8}$ vs.\ $2.5\times10^{10}$). In the small set, the argon program shows a $\frac{1}{46}$ ratio ($4.5M$ vs.\ $207M$).
This result demonstrates \qsndl's compilation performance, especially on larger real-world quantum applications.

The gate reduction barely changes with the error bound: it stays nearly the same from $0.1$ down to $0.0002$.
It grows slightly with the time period, from about $94\%$ at $\frac{\pi}{16}$ to $96$--$97\%$ at $\pi$ and $4\pi$.

In comparing \qsndl with PennyLane, we found an issue in PennyLane's gate estimator.
PennyLane's gate estimator compares quantum gates without decomposition; e.g., they do not decompose long-chain control gates, such as $4$-qubit Toffoli gates, into $ 1$- and $ 2$-qubit gates, while the \qsndl estimator decomposes everything into $ 1$- and $ 2$-qubit gates. PennyLane's gate estimator may not reflect the real cost of quantum gates, as some long-chain control gates might be costly to implement. Therefore, we utilize the gate estimator in \qsndl for the comparison.

\subsection{RQ3: Correctness and Reliability}\label{sec:correct-comparison}

\ignore{
\begin{figure}[t]
\begin{minipage}[b]{1\linewidth}
    \centering
    \includegraphics[width=1\linewidth]{fig13_overall_qblue_steps_pi16.png}
    \end{minipage}
\begin{minipage}[b]{1\linewidth}
    \centering
    \includegraphics[width=1\linewidth]{fig13_overall_qblue_steps_pi4.png}
    \end{minipage}
\caption{QBlue optimization reduction compared to other platforms for error bounds:  $0.0002$, $0.002$, $0.02$, and $0.1$, and simulation time: $4\pi$, $\pi$, $\frac{\pi}{4}$ and $\frac{\pi}{16}$. In each figure, the upper panel compares against PennyLane.}
\label{tab:splitting-step1}
\end{figure}
}

\begin{figure}[t]
{\footnotesize
    \centering
    \includegraphics[width=1\linewidth]{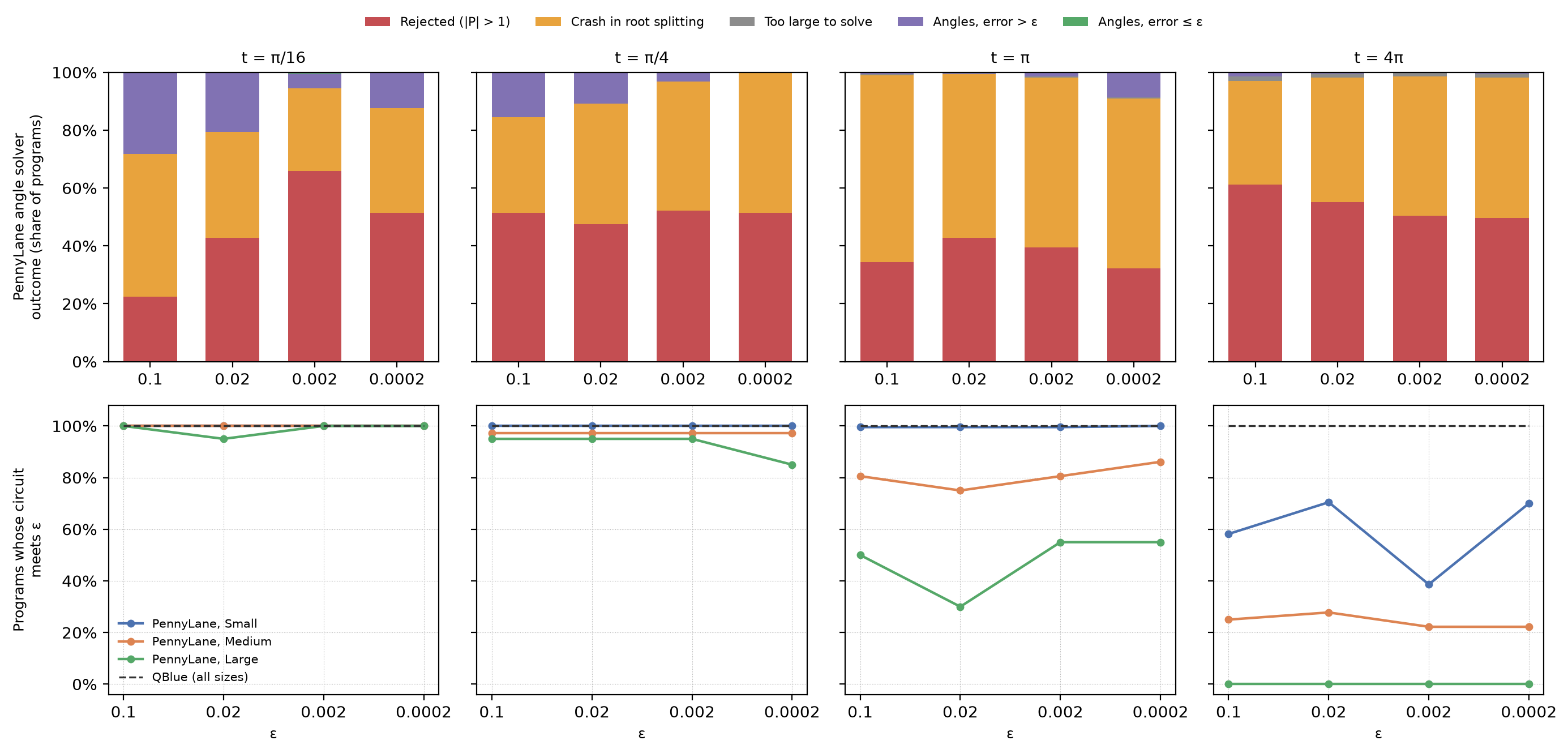}
}
\caption{PennyLane's correctness and reliability for error bounds $0.1$, $0.02$, $0.002$, and $0.0002$, and time periods $\frac{\pi}{16}$, $\frac{\pi}{4}$, $\pi$, and $4\pi$, over all $276$ programs. Top: the outcome of PennyLane's angle solver. Bottom: the share of programs whose PennyLane circuit meets the requested error bound; \qsndl meets it in every case.}
\label{fig:pl-reliability}
\end{figure}

We now examine the correctness and reliability of \qsnd by comparing it with PennyLane.
Correctness measures whether a compiler has errors in the produced circuits, while reliability measures whether a compiler fails to produce circuits. We document the issues in PennyLane in \Cref{fig:pl-reliability}.

PennyLane has a reliability issue for the QSP algorithm. In QSP, the compiler is supposed to generate the right angles to approximate the correct circuits; i.e., the algorithm includes an angle solver. If the solver does not produce the correct angles, the compiler fails to produce the correct circuits.
The angle solver in \qsnd is verified to be correct and handles all cases correctly.

The main issue with PennyLane's angle solver is that it relies on generating polynomial degrees and checks only one Bessel term.
We ran PennyLane's angle solver on all $4{,}416$ cases ($276$ programs across $4$ error bounds and $4$ time periods), and the results are in the top row of \Cref{fig:pl-reliability}. The solver produced usable angles in only one case.
In $2{,}071$ cases it rejected the polynomial because its value exceeded $1$, and in $2{,}022$ cases it crashed in a key step, root splitting. In $20$ cases, the polynomial was too large to solve; in the remaining $302$ cases, it returned angles whose error exceeded the user's bound.

One key factor in the correctness guarantees is how it handles error bounds.
When users input an error bound, \qsnd correctly produces a circuit to meet the error bound requirement. This is not the case in PennyLane.
As shown in the bottom row of \Cref{fig:pl-reliability}, PennyLane meets the error bound on only $191$ of the $320$ large-program cases (large programs acorss error bounds and $4$ time periods), and on none of them at $4\pi$ \footnote{We utilize the \qsnd angle solver for conducting the experiment since PennyLane's angle solver fails to produce results.}. For the other programs, PennyLane generates incorrect circuits with an error bound larger than the user-required one. In the worst cases, the error is about $10^4$ times the requested bound, which means the circuit does not approximate $\eexp{\uapp{r}{\hat{H}}}$ at all.
Overall, in PennyLane, there are $3{,}785$ cases where the produced circuits meet the error bound requirement, $611$ cases producing circuits with larger errors than the user-defined ones, and there are $20$ cases when PennyLane fails to handle the error bound, i.e., it runs forever to find the correct circuit.

\section{A Thorough Study on Comparing BE and Trotter-Based Algorithms}
\label{sec:discussion}

The \qsnd compiler includes BE- and Trotter-based algorithms for compiling quantum simulation programs, and it lets users set up the target compilation machines, either digital- or analog-based.
Among these different machine choices, algorithms, and circuit-level optimizations, which selection is likely to produce the best compilation result for a program, mainly in terms of resource estimation? This section compares the different possible compilation paths in \qsnd and answers the question.

\begin{figure}[t]
{\footnotesize
    \centering
    \includegraphics[width=1\linewidth]{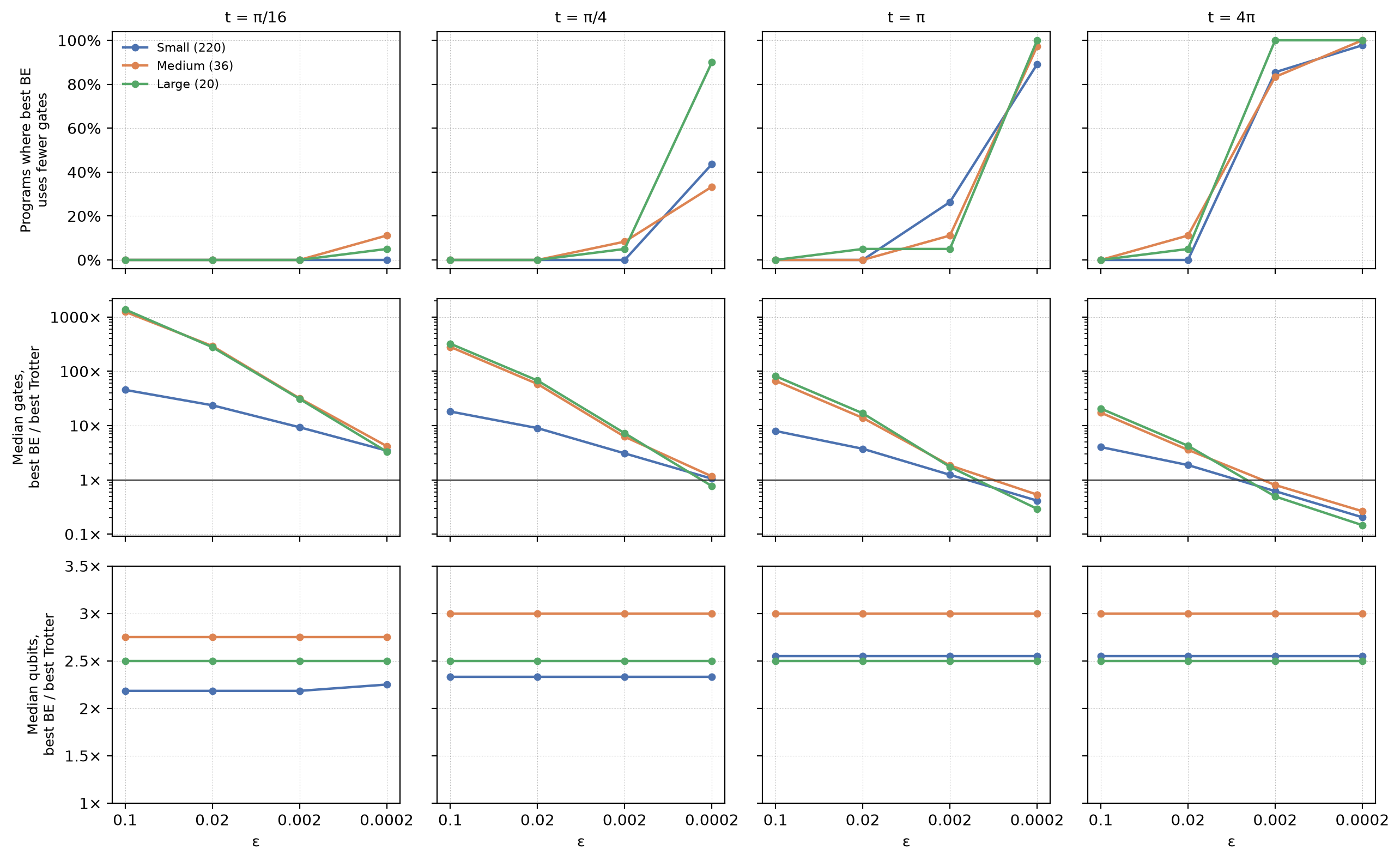}
}
\caption{Comparing the best-performing Trotter-based and BE-based algorithms; error bounds: $0.0002$, $0.002$, $0.02$, and $0.1$, and simulation time periods: $\pi/16$, $\pi/4$, $\pi$, $4\pi$. The first row records the percentage of BE-based algorithms that perform better; the second row measures the gate multiples that the BE-based algorithms spent more than Trotter-based ones; and the third row shows the qubit multiples that the BE-based algorithms used more.}
\label{tab:trotter-selecta}
\end{figure}

\subsection{Comparing BE-based and Trotter-based Algorithms}\label{sec:diff-trotter}

We first compare the BE-based algorithms with the Trotter-based ones, as there is a long-standing argument about which algorithm produces the most efficient code.
When comparing the two, we set up different error bounds and simulation time periods. For each error bound and time period, we run the \qsnd compiler once for the BE-based algorithms and multiple times for the Trotterization-based algorithms, executing every program in \Cref{tab:data-table}.
We then pick the best BE-based algorithm and compare it with the best Trotterization algorithm for each program.
The result is summarized in \Cref{tab:trotter-selecta}.

Overall, Trotter-based algorithms save more qubits than the BE-based algorithms, because BE-based compilation generates ancilla qubits. 
On average, BE-based algorithms require 2-3 times as many qubits as the best Trotterization-based algorithms.

In terms of gate savings, we got a surprising result: BE-based algorithms should produce a much more efficient circuit. 
However, this is not always true. For most error-bound and time-period pairs, the Trotter-based algorithm, usually second-order Trotterization, produces the most efficient circuit. 
For time periods up to $\pi$ and error bounds of $0.002$ or looser, Trotter-based algorithms save more gates than BE-based algorithms.
For the $0.002$ error bound, Trotter-based algorithms save more than $90\%$ of the gates in many cases, especially for the time periods less than $\pi/4$.
For the time period $\pi/16$ and the error bound $0.1$, the Trotter-based algorithm saves more than $99.9\%$ of the gates.
Especially, Trotter-based algorithms tend to save more qubits for medium and large programs.
BE-based algorithms gain ground as the time period grows and the error bound tightens: they win on most programs at $\pi$ and $4\pi$ with $\epsilon=0.0002$, and at $4\pi$ with $\epsilon=0.002$.
The issue is that most real-world quantum simulation tasks do not require long simulation times, e.g., they typically require less than a second, so the ability to handle longer simulation periods is not very attractive to users. In most quantum chemistry applications, the error bound up to $0.002$ is enough, whereas quantum physics applications require a high-precision error bound ($0.0002$).
As we can see, the BE-based algorithms will be useful for high-precision applications with enough ancilla qubits, while Trotter-based algorithms will be useful for many other applications.

\subsection{The Effects of Different BE-based Algorithms}\label{sec:diff-lcu}
\begin{table}[t]
{\footnotesize
\begin{tabular}{ll|rrr|rrr|rrrr}
\toprule
 & & \multicolumn{3}{c|}{QSVT is best} & \multicolumn{3}{c|}{QSP is best} & \multicolumn{4}{c}{Gates / best BE (median)} \\
$t$ & $\epsilon$ & S & M & L & S & M & L & QSVT & QSP & Taylor & Walk \\
\midrule
$\pi/16$ & 0.1 & 2 & 14 & 20 & 218 & 22 & 0 & 1.2$\times$ & 1.0$\times$ & 3.8$\times$ & 2.2$\times$ \\
$\pi/16$ & 0.02 & 3 & 14 & 20 & 217 & 22 & 0 & 1.1$\times$ & 1.0$\times$ & 4.0$\times$ & 4.3$\times$ \\
$\pi/16$ & 0.002 & 5 & 14 & 20 & 215 & 22 & 0 & 1.1$\times$ & 1.0$\times$ & 4.3$\times$ & 3.5$\times$ \\
$\pi/16$ & 0.0002 & 25 & 14 & 20 & 195 & 22 & 0 & 1.1$\times$ & 1.0$\times$ & 4.1$\times$ & 3.0$\times$ \\
\midrule
$\pi/4$ & 0.1 & 143 & 34 & 20 & 77 & 2 & 0 & 1.0$\times$ & 1.2$\times$ & 5.9$\times$ & 6.0$\times$ \\
$\pi/4$ & 0.02 & 139 & 34 & 20 & 81 & 2 & 0 & 1.0$\times$ & 1.2$\times$ & 5.4$\times$ & 5.5$\times$ \\
$\pi/4$ & 0.002 & 140 & 34 & 20 & 80 & 2 & 0 & 1.0$\times$ & 1.2$\times$ & 6.0$\times$ & 5.0$\times$ \\
$\pi/4$ & 0.0002 & 139 & 34 & 20 & 81 & 2 & 0 & 1.0$\times$ & 1.3$\times$ & 6.6$\times$ & 4.7$\times$ \\
\midrule
$\pi$ & 0.1 & 219 & 36 & 20 & 1 & 0 & 0 & 1.0$\times$ & 1.4$\times$ & 6.9$\times$ & 6.7$\times$ \\
$\pi$ & 0.02 & 220 & 36 & 20 & 0 & 0 & 0 & 1.0$\times$ & 1.5$\times$ & 7.8$\times$ & 6.4$\times$ \\
$\pi$ & 0.002 & 219 & 36 & 20 & 1 & 0 & 0 & 1.0$\times$ & 1.6$\times$ & 8.7$\times$ & 6.2$\times$ \\
$\pi$ & 0.0002 & 219 & 36 & 20 & 1 & 0 & 0 & 1.0$\times$ & 1.6$\times$ & 9.3$\times$ & 5.8$\times$ \\
\midrule
$4\pi$ & 0.1 & 220 & 36 & 20 & 0 & 0 & 0 & 1.0$\times$ & 1.5$\times$ & 8.0$\times$ & 6.6$\times$ \\
$4\pi$ & 0.02 & 220 & 36 & 20 & 0 & 0 & 0 & 1.0$\times$ & 1.5$\times$ & 8.7$\times$ & 6.3$\times$ \\
$4\pi$ & 0.002 & 220 & 36 & 20 & 0 & 0 & 0 & 1.0$\times$ & 1.6$\times$ & 9.2$\times$ & 6.0$\times$ \\
$4\pi$ & 0.0002 & 220 & 36 & 20 & 0 & 0 & 0 & 1.0$\times$ & 1.7$\times$ & 9.7$\times$ & 6.3$\times$ \\
\bottomrule
\end{tabular} }
\caption{Comparing the four BE-based algorithms. Left: the number of programs where QSVT or QSP produces the fewest gates, per dataset (S: $220$, M: $36$, L: $20$ programs); Taylor series and quantum walk are never the best. Right: each algorithm's gate count divided by the best BE-based algorithm's, median over all $276$ programs.}
\label{tab:be-algorithms}
\end{table}

\ignore{
\begin{table*}[t]
\centering
{\tiny
\setlength{\tabcolsep}{4pt}
\begin{tabular}{llrrrrrrrrrrrr}
\toprule
Size & Metric & \multicolumn{2}{c}{$\varepsilon=0.0002$} & \multicolumn{2}{c}{$\varepsilon=0.002$} & \multicolumn{4}{c}{$\varepsilon=0.02$} & \multicolumn{4}{c}{$\varepsilon=0.1$} \\
\cmidrule(lr){3-4} \cmidrule(lr){5-6} \cmidrule(lr){7-10} \cmidrule(lr){11-14}
& & $\pi$ & $4\pi$ & $\pi$ & $4\pi$ & $\pi/16$ & $\pi/4$ & $\pi$ & $4\pi$ & $\pi/16$ & $\pi/4$ & $\pi$ & $4\pi$ \\
\midrule
Small   & QSVT wins      & 219.0 & 220.0 & 219.0 & 220.0 &   3.0 & 139.0 & 220.0 & 220.0 &   2.0 & 143.0 & 219.0 & 220.0 \\
Small   & QSP wins       &   1.0 &   0.0 &   1.0 &   0.0 & 217.0 &  81.0 &   0.0 &   0.0 & 218.0 &  77.0 &   1.0 &   0.0 \\
Small   & QSVT sav. (\%) & 89.2\% & 89.9\% & 88.4\% & 89.1\% & 66.6\% & 82.2\% & 87.1\% & 88.2\% & 61.9\% & 81.0\% & 85.6\% & 87.4\% \\
Small   & QSP sav. (\%)  & 82.9\% & 83.2\% & 82.1\% & 82.6\% & 71.3\% & 79.9\% & 81.3\% & 82.0\% & 69.8\% & 78.9\% & 79.8\% & 81.3\% \\
\midrule
Medium  & QSVT wins      &  36.0 &  36.0 &  36.0 &  36.0 &  14.0 &  34.0 &  36.0 &  36.0 &  14.0 &  34.0 &  36.0 &  36.0 \\
Medium  & QSP wins       &   0.0 &   0.0 &   0.0 &   0.0 &  22.0 &   2.0 &   0.0 &   0.0 &  22.0 &   2.0 &   0.0 &   0.0 \\
Medium  & QSVT sav. (\%) & 88.9\% & 89.7\% & 88.1\% & 89.0\% & 76.4\% & 85.4\% & 86.9\% & 88.1\% & 73.9\% & 84.0\% & 85.9\% & 87.2\% \\
Medium  & QSP sav. (\%)  & 82.7\% & 83.0\% & 82.0\% & 82.6\% & 76.8\% & 79.9\% & 80.9\% & 81.9\% & 75.7\% & 79.1\% & 80.4\% & 81.3\% \\
\midrule
Large   & QSVT wins      &  20.0 &  20.0 &  20.0 &  20.0 &  20.0 &  20.0 &  20.0 &  20.0 &  20.0 &  20.0 &  20.0 &  20.0 \\
Large   & QSP wins       &   0.0 &   0.0 &   0.0 &   0.0 &   0.0 &   0.0 &   0.0 &   0.0 &   0.0 &   0.0 &   0.0 &   0.0 \\
Large   & QSVT sav. (\%) & 89.3\% & 89.8\% & 88.6\% & 88.9\% & 84.1\% & 87.8\% & 87.9\% & 88.1\% & 82.9\% & 87.0\% & 86.5\% & 87.5\% \\
Large   & QSP sav. (\%)  & 82.9\% & 82.9\% & 82.3\% & 82.2\% & 79.4\% & 80.4\% & 81.9\% & 81.6\% & 78.4\% & 79.9\% & 80.8\% & 81.5\% \\
\midrule
Tallied & QSVT wins      & 275.0 & 276.0 & 275.0 & 276.0 &  37.0 & 193.0 & 276.0 & 276.0 &  36.0 & 197.0 & 275.0 & 276.0 \\
Tallied & QSP wins       &   1.0 &   0.0 &   1.0 &   0.0 & 239.0 &  83.0 &   0.0 &   0.0 & 240.0 &  79.0 &   1.0 &   0.0 \\
Tallied & QSVT sav. (\%) & 89.1\% & 89.8\% & 88.4\% & 89.0\% & 75.7\% & 85.1\% & 87.3\% & 88.1\% & 72.9\% & 84.0\% & 86.0\% & 87.4\% \\
Tallied & QSP sav. (\%)  & 82.8\% & 83.0\% & 82.1\% & 82.5\% & 75.8\% & 80.0\% & 81.4\% & 81.9\% & 74.6\% & 79.3\% & 80.4\% & 81.4\% \\
\bottomrule
\end{tabular}
}
\caption{Comparison of QSVT and QSP compilation performance across Small (220 programs), Medium (36 programs), and Large (20 programs) sets, with tallied totals across sizes, for error tolerances $\varepsilon \in \{0.0002, 0.002, 0.02, 0.1\}$ and evolution times $t \in \{\pi/16, \pi/4, \pi, 4\pi\}$. Win counts report how often each method achieves the lowest gate count among LCU-based algorithms (ties split evenly; Taylor series and Quantum walk record $0.0$ wins). Percentage rows report average gate reduction relative to the \texttt{Taylor series} baseline (positive values indicate fewer gates than \texttt{Taylor series}).}
\label{tab:lcu-qsvt-qsp}
\end{table*}
}

We now compare the four BE-based algorithms in \qsnd against each other. For each program, error bound, and time period, we record which algorithm produces the fewest gates and how far behind the others are, as demonstrated in \Cref{tab:be-algorithms}.

For each program, we first identify the algorithm that produces the optimal circuit, and we find that either QSVT or QSP wins, while the Taylor series and quantum walk algorithms never win.
Additionally, we find that the time period helps determine which of the two algorithms, QSVT and QSP, performs better.
At time period $\frac{\pi}{16}$, QSP produces optimal circuits on almost all small programs and on $22$ of the $36$ medium programs, while QSVT produces optimal circuits for all large programs.
From $\frac{\pi}{4}$ on, QSVT takes over: it produces optimal circuits for $140$ out of the $220$ small programs, and for all but at most one program at the time periods $\pi$ and $4\pi$.
However, the difference in gate counts between the two algorithms is relatively small; typically, one algorithm requires $6$--$37\%$ fewer gates.
QSP uses at most $1.7$ times as many gates as QSVT for median-set programs, while it saves $10\%$-$20\%$ gates for the $\frac{\pi}{16}$ time-period case.

The Taylor series and quantum walk algorithms are much further behind. They need $2$--$10$ times more gates than the best algorithm, and the gap grows with the time period, so the best algorithm saves $74$--$89\%$ of the gates over the worst one.
Qubit counts differ too: QSP uses the fewest qubits, QSVT and quantum walk need $4$--$5$ more ancillas, and the Taylor series needs $2$--$5$ times as many qubits as QSP, because each Taylor term needs its own ancilla register.

\subsection{Comparison against Analog-based Machines}

\begin{table}[t]
\centering
{\footnotesize
\begin{tabular}{llrrrrrr}
\toprule
 & & \multicolumn{3}{c}{1-qubit gate count reduction (analog)} & \multicolumn{3}{c}{Multi-qubit gate count reduction (analog)} \\
\cmidrule(lr){3-5}\cmidrule(lr){6-8}
$t$ & $\epsilon$ & S & M & L & S & M & L \\
\midrule
$\pi/16$ & 0.1 & 91\% & 99\% & 99\% & 99\% & 99\% & 99\% \\
$\pi/16$ & 0.02 & 83\% & 98\% & 98\% & 99\% & 99\% & 99\% \\
$\pi/16$ & 0.002 & 57\% & 81\% & 81\% & 97\% & 99\% & 99\% \\
$\pi/16$ & 0.0002 & -23\% & -32\% & -75\% & 92\% & 93\% & 94\% \\
\midrule
$\pi/4$ & 0.1 & 78\% & 98\% & 98\% & 99\% & 99\% & 99\% \\
$\pi/4$ & 0.02 & 55\% & 89\% & 92\% & 97\% & 99\% & 99\% \\
$\pi/4$ & 0.002 & -33\% & -1\% & 21\% & 92\% & 95\% & 97\% \\
$\pi/4$ & 0.0002 & -296\% & -409\% & -549\% & 75\% & 73\% & 78\% \\
\midrule
$\pi$ & 0.1 & 52\% & 91\% & 93\% & 97\% & 99\% & 99\% \\
$\pi$ & 0.02 & -3\% & 55\% & 67\% & 94\% & 98\% & 99\% \\
$\pi$ & 0.002 & -209\% & -220\% & -220\% & 81\% & 84\% & 88\% \\
$\pi$ & 0.0002 & -838\% & -950\% & -1221\% & 43\% & 44\% & 55\% \\
\midrule
$4\pi$ & 0.1 & -1\% & 63\% & 73\% & 94\% & 98\% & 99\% \\
$4\pi$ & 0.02 & -120\% & -70\% & -32\% & 86\% & 92\% & 95\% \\
$4\pi$ & 0.002 & -536\% & -595\% & -774\% & 60\% & 62\% & 70\% \\
$4\pi$ & 0.0002 & -1927\% & -2028\% & -2542\% & -15\% & -14\% & 7\% \\
\bottomrule
\end{tabular}
}
\caption{Relative reduction in 1-qubit and multi-qubit gate counts for the best Trotter compilation on the Indiana analog backend compared against the best block-encoding (BE) circuit on the IBM digital backend across Small (S), Medium (M), and Large (L) programs for evolution times $t \in \{\pi/16, \pi/4, \pi, 4\pi\}$ and error tolerances $\epsilon \in \{0.1, 0.02, 0.002, 0.0002\}$. Each entry compares the median counts per dataset; for the analog backend, gates are pulses. Positive values indicate that the analog compilation requires that percentage fewer gates than the digital BE circuit; negative values indicate more, e.g., $-100\%$ means twice as many.}
\label{tab:gate-reduction-analog-vs-be}
\end{table}

One interesting comparison is to examine the power of analog-based quantum machines.
As demonstrated above, BE-based algorithms directly generate digital quantum gates, while Trotter-based algorithms generate small unit simulation operations based on Pauli strings.
This means analog-based quantum machines might support only Trotter-based ones.
This is essentially another way to compare BE-based and Trotterization-based approaches: comparing digital circuit generation by BE-based algorithms with analog pulse generation via Trotter-based algorithms.
Below, we compare gate counts, while qubit counts match those documented in \Cref{tab:trotter-selecta}.

Several studies have used analog-based quantum computers for Hamiltonian simulation, believing they can deliver better performance \cite{10.1145/3632923}.
Using \qsnd, we examine this claim by comparing resource usage for compiling Hamiltonian simulation programs to digital and analog target machines in \Cref{tab:gate-reduction-analog-vs-be}.
Gates on a digital machine and pulses on an analog machine run at different speeds \cite{Kyprianidis_2024,Alexander_2020}; here we compare their raw counts, i.e., how many operations each machine has to execute.

The actual analog machine prototype is based on the Indiana analog machine, a new kind of analog quantum computer that permits high-dimensional interactions.
\Cref{tab:gate-reduction-analog-vs-be} compares circuit compilation for the Indiana analog machine with the digital compilation to the IBM machine.
The digital circuits are optimized by VOQC.
While the results for $1$-qubit terms may be worse, the Indiana analog backend significantly reduces operations for multi-qubit terms.
For time periods up to $\frac{\pi}{4}$, it needs at least $73\%$ fewer multi-qubit operations on every dataset and error bound, and $99\%$ fewer at the error bounds $0.1$ and $0.02$.
The advantage shrinks as the time grows and the error bound tightens: at the time period $\pi$ with the error bound $\epsilon=0.0002$, the reduction is $43$--$55\%$, and at $4\pi$ with $\epsilon=0.002$ it is $60$--$70\%$.
Even for the case of a $4\pi$ time period and $0.0002$ error bound \footnote{The case of $4\pi$ and $0.0002$ error bound is rarely used, because the $0.0002$ error bound is usually used for quantum physics applications which do not require a long simulation time for $4\pi$.}, the Indiana prototype performs worse on small-set programs, while it reduces operations on large-set programs.
This is because the Indiana analog backend can execute a multi-qubit term at once, whereas many Hamiltonian simulation programs contain such long-range multi-qubit terms.

In quantum computer executions, saving multi-qubit terms matters much more than saving $1$-qubit terms, since multi-qubit gates are the main source of performance slowdown and noise. Additionally, the Indiana analog backend does not adopt typical one-qubit term optimization, such as merging sequences of one-qubit terms into a single $SU(2)$ term \cite{Atas_2021}. Indeed, the above experiment shows the power of gate-level optimization \cite{VOQC}, while analog-based compilation lacks such optimization, suggesting that studying it could improve analog-based compilation for Hamiltonian simulations. Specifically, the Indiana analog backend might edge out digital approaches for Hamiltonian simulation, but sophisticated analog-based optimizations are needed to realize this advantage.
 \section{Related Work}
\label{sec:related}

This section reviews the most relevant related work to our approach. While many other contributions exist, we focus on those directly connected to the Hamiltonian simulation and compilation techniques discussed in this paper.

\noindent{\textbf{\emph{Quantum Circuit-based Programming Languages.}}}
Quantum circuit-based language development is a central focus in the field. Several industrial quantum circuit languages have been developed, including Q\#~\cite{Svore_2018}, Quilc~\cite{smith2020opensource}, ScaffCC~\cite{JavadiAbhari_2015}, ProjectQ~\cite{Steiger_2018}, Cirq~\cite{cirq_developers_2023_10247207}, and Qiskit~\cite{Qiskit2019}. While these languages lack formal semantics, they are grounded in a standardized denotational semantics for quantum circuits. In contrast, a number of tools focus on formal verification of quantum circuit programs. These include \qwire~\cite{RandThesis}, \sqir~\cite{PQPC}, and \qbricks~\cite{qbricks}, as well as quantum Hoare logic and its extensions~\cite{qhoare,qhoreusage,10.1145/3571222}, and Qafny~\cite{li2024qafny}. These systems have been used to verify a range of quantum algorithms, including Grover’s search and quantum phase estimation. Verification has also been applied to quantum circuit optimizations, such as in \voqc~\cite{VOQC} and CertiQ~\cite{Shi2019}, and to quantum circuit compilation processes, as demonstrated by ReVerC~\cite{reverC} and ReQWIRE~\cite{Rand2018ReQWIRERA}.

Several functional languages have been developed to model quantum circuits with formal semantics, including~\cite{10.1007/978-3-319-89366-2_19,DIAZCARO2019104012,Selinger2013,van_Tonder_2004,Altenkirch,10.1007/978-3-642-40922-6_5,Voichick_2023,Green2013,mingsheng-ying,qml-update,qml-thesis,10.1145/3571204}. These languages are typically designed to encode quantum circuits as functional constructs, such as through quantum pattern matching. Silq~\cite{sliqlanguage} introduces automatic uncomputation within a denotational semantics framework for describing circuit behavior. Finally, the ZX-calculus~\cite{Coecke_2011,wang2023completeness} provides a diagrammatic language for expressing unitary operations without measurement, offering an alternative to algebraic quantum semantics.

\noindent{\textbf{\emph{Analog Quantum Computing Languages and Pulse Level Programming.}}}
Several pulse-level and target-machine Hamiltonian programming interfaces treat quantum computers as analog quantum simulators. Notable examples include IBM Qiskit Pulse~\cite{10.1145/3505636}, QuEra Bloqade~\cite{bloqade2023quera}, and Pasqal Pulser~\cite{Silverio2022pulseropensource}, all developed by hardware service providers. In parallel, computational quantum physics packages such as QuTiP~\cite{JOHANSSON20121760} support Hamiltonian simulation on classical computers. These efforts are inspired by a long history of classical analog compilation techniques~\cite{10.1145/3373376.3378449,10.1145/2980983.2908116}. SimuQ~\cite{10.1145/3632923} introduces a formal language for analog quantum computing based on the construction of Hamiltonians using Pauli groups. Recent work by Cao \emph{et al.} \cite{cao2024marqsimreconcilingdeterminismrandomness} proposes a Trotterization algorithm that improves upon QDrift.

Compared to these approaches, \qsnd provides the first verified compilation framework that transforms user-level Hamiltonian programs written in second-quantization form into both digital and analog quantum circuits.

\noindent{\textbf{\emph{Quantum Simulation related Quantum computational Approaches.}}}
Many quantum computation software and algorithms have been proposed for analyzing particle systems.
There are previous software tools \cite{10.1145/3511715,POWERS2021100696,li2025verifiedcompilerquantumsimulation} and algorithms \cite{10.1145/3503222.3507715,PhysRevA.109.042418,10.1145/3729269,Campbell_2019} for improving Trotter-based Hamiltonian simulation compilation algorithms. Particularly, \qsnd \cite{li2025verifiedcompilerquantumsimulation} develop a verified compiler for Trotter-based algorithms.
There are BE-based compilation algorithms for Hamiltonian simulations \cite{Andrew2018,Low_2019,Berry_2015b,Low_2017,Martyn_2021}, where \qsndl examines all these algorithms by verifying them in the compilation pipeline and conducting an empirical study. 

Many other works \cite{Cervera_Lierta_2018,Yang_2020,YAMAGUCHI2002343,doi:10.1021/acs.jctc.2c00974} discuss algorithms simulating different particle systems in quantum computers. Many algorithms are not implementable in current quantum computer architectures due to machine limitations.
The variational quantum eigensolver is one of the most well-known algorithms \cite{Tilly_2022,Cerezo2022} to analyze particle systems.
It is mainly used for computing energy states, such as estimating the ground-state energy of water molecules \cite{Nam2020}.

Some industrial tools are available for the computation of particle systems, including OpenFermion \cite{mcclean2019openfermionelectronicstructurepackage}, Qiskit Nature \cite{qiskit2024}, and PennyLane \cite{bergholm2022pennylaneautomaticdifferentiationhybrid}. These tools are mostly used to permit the computation of energy state computation, not necessarily focusing on Hamiltonian simulation.
PennyLane is the only tool having compilation pipeline based BE-based algorithms.
 \section{Conclusion and Future Work}\label{sec:conclusion}

This paper extends \qsnd with BE-based compilation algorithms, named \qsndl, and empirically evaluates them by comparing \qsndl with Trotter-based algorithms.
We verify the entire compilation pipeline in \qsndl in \rocq; that is, we fully verify the compilation pipeline for Hamiltonian simulation programs across different quantum computer backends and provide a comprehensive evaluation of these backends under various settings, including error bounds, simulation time, and dataset sizes.

Our empirical study shows that \qsndl clearly outperforms the previous BE-based compiler, PennyLane, in efficiency and correctness. 
We discover several issues and bugs in PennyLane. In particular, we find that PennyLane's angle solver cannot find the correct angles for the QSP algorithm, and PennyLane fails to compile many programs with large Pauli strings.

We also find a surprising result: BE-based algorithms may not offer an efficient advantage over the best Trotter-based algorithms for most time periods and error-bound settings, even though they have a theoretical advantage.
In addition, Trotter-based algorithms are implementable on analog quantum computers, while BE-based algorithms are not, and Trotter-based algorithms implemented on an analog machine show an even better efficiency advantage over BE-based algorithms.

For future work, we plan to develop tools to analyze Hamiltonian-based programs and integrate the \qsnd compiler with ground-state energy solvers for scientific computation.
One previous framework \cite{10.1145/3747521} that performs the task and naturally extends Hamiltonian simulation with adiabatic quantum computation (AQC), i.e., 5he \qsnd application semantics can be naturally extended to AQC for energy state computation.
 
\bibliography{reference}

\newpage
\appendix
\section{Syntax and Semantics for Pauli String based Hamiltonian}\label{sec:formal}

Before we discuss the \qsnd compilation pipeline, we examine its formal semantics.
This section discusses \qsnd's syntax and semantics within the Pauli string formalism.

\myparagraph{Syntax.}
\Cref{fig:paulisyntax} defines the \qsnd program syntax. A constraint expression $\hat{H}$ denotes a symbolic Hamiltonian imposing user-defined constraints to a system, while a unitary expression $U=\eexp{\hat{H}}$ represents compiled time evolution, executable on quantum hardware.

\begin{figure}[h]
{\small
  \[\begin{array}{l@{\quad}l@{\;\;}c@{\;\;}l@{\;\;}} 
\text{Pauli Terms} & \pau  & ::= & X \mid Y \mid Z \mid I
\\[0.2em]
       \text{Constraint Expression} & \hat{H} & ::= & \iseq{r}{\pau} \mid \hat{H} + \hat{H} \mid \hat{H} \otimes \hat{H} 
       \\[0.2em]
       \text{Unitary Expression} & U & ::= & \eexp{\hat{H}} \mid \sapp{U}{U}
    \end{array}
  \]
}
  \caption{\qsnd Pauli-string-based syntax.} 
  \label{fig:paulisyntax}
\end{figure}

The term $\iseq{r}{\pau}$ represents a single Pauli operation ($\{X,Y,Z, I\}$), scaled by a real amplitude $r$, which is enough to represent any Hermitian Hamiltonian; see \Cref{sec:compile-particlea}. 
Other expressions include tensors $e \otimes e$ to represent operations on distinct qubits and additive combinations $e + e$ for linear combinations of terms.
At the unitary level, $\eexp{\hat{H}}$ performs Hamiltonian simulation, and $\sapp{U}{U}$ represents unitary composition. 
Programs often use indexed notation as syntactic sugar for tensor expressions. For a system with $n$ qubits, an indexed operator such as $\pau(j)$ denotes the tensor product of identity operators at all qubits except for $j$, where the operator $\pau$ is applied. For instance, the term $\frac{1}{2}(X(0) \circ  X(1) + Y(0) \circ Y(1))$ refers to $(X \otimes I) \circ (I \otimes X) + (Y \otimes I) \circ (I \otimes Y) = X \otimes X + Y \otimes Y$.

\begin{figure*}[t]
{\small
{
\begin{center}
$\hspace{-0.5em}
\denote{X} := \begin{bmatrix}
0 & 1\\
1 & 0
\end{bmatrix}
\quad\;\;
\denote{Y} := \begin{bmatrix}
0 & \sminus i\\
i & 0
\end{bmatrix}
\quad\;\;
\denote{Z} := \begin{bmatrix}
1 & 0\\
0 & \sminus 1
\end{bmatrix}
\quad\;\;
\denote{I} := 
\begin{bmatrix}
1 & 0\\
0 & 1
\end{bmatrix}
$
\\[0.7em]
\end{center}
}
}
{\small 
       \begin{mathpar}
        \inferrule[]{}
       {\denote{r \pau} := r \denote{\pau}}
       
        \inferrule[]{}
        { \denote{\hat{H} \otimes \hat{H}'} := \denote{\hat{H}} \,\textcolor{purple}{\otimes}\, \denote{\hat{H}'} }

        \inferrule[]{}
        { \denote{\hat{H} + \hat{H}'} := \denote{\hat{H}} \,\textcolor{purple}{+}\, \denote{\hat{H}'} }

    \inferrule[]{}
        {\denote{\eexp{\hat{H}}} := \textcolor{purple}{\sum_{n=0}^{\infty}}\frac{\textcolor{purple}{(}\sminus i \denote{\hat{H}}_g\textcolor{purple}{)^n}}{n!} }
                
        \inferrule[]{}
        { \denote{\sapp{U}{U'}} := \denote{U}(\denote{U'}) }
  \end{mathpar}
}
\caption{\qsnd semantics for Pauli-string based Hamiltonians and unitaries. Terms inside $\denote{-}$ are \qsnd operators. \textcolor{purple}{purple} items are quantum state vector operations, and $\denote{-}(\denote{-})$ is matrix multiplication.} \label{fig:sem}
\end{figure*}

\myparagraph{Semantics.}
The \qsnd programs describe finite $n$-qubit systems, with the denotational semantics in in \Cref{fig:sem}, defined as $\denote{\hat{H}}$ for interpreting Hamiltonian constraints and $\denote{U}$ for interpreting unitary operations. 
The semantics essentially interprets the $n$-qubit system as a $2^n \times 2^n$ matrix.

The first two lines in \Cref{fig:sem} define the Hamiltonian constraint semantics, where the first line interprets each Pauli operator as its matrix form.
In the second line, we interpret the tensor and sum operation as a matrix tensor ($\textcolor{purple}{\otimes}$) and a linear sum ($\textcolor{purple}{+}$), respectively.

The last line defines the \qsnd dynamic semantics. The left one interprets the expression $\eexp{\hat{H}}$ as the matrix exponential, provided that $\hat{H}$ is Hermitian. The exponential is defined via its standard power series expansion, which converges to a unitary operator when $\hat{H}$ is Hermitian. We interpret this construct as the limit of a convergent series of the matrix exponential series, approximated computationally as needed.
The right interprets sequential composition of unitaries as standard matrix multiplication. Thus, dynamic programs in \qsnd are built by composing unitary evolutions, each derived from a Hamiltonian constraint via simulation.

\section{The QBlue Trotter-based Compilation Procedure}\label{sec:compilationa}

This section describes the \qsnd compilation steps and explains how \rocq structures and verifies their correctness proofs.
The compiler transforms a Hamiltonian simulation expression $\eexp{\uapp{r}{\hat{H}}}$, where $e$ is a \qsnd Hamiltonian program and $r$ is a time parameter, into a quantum circuit, approximating $\eexp{\uapp{r}{\hat{H}}}$ with a user-defined error bound $\epsilon$, executable on quantum hardware. 
We model compilation as a judgment of the form $(\hat{H}, r) \gg U$. The \qsnd compiler supports multiple compilation paths, letting users choose from four Trotterization algorithms and three circuit backends. 
We first show the correctness theorem for compiling a Hamiltonian simulation of a constraint expression $e$ to quantum circuits.

\begin{theorem}[Trotter-based Algorithm Compilation Correctness]\label{thm:compile-gooda}\rm 
Given $\hat{H}$ in \Cref{fig:paulisyntax}, a time period $r$, and a desired error $\epsilon$, we compile its simulation $\eexp{\uapp{r}{\hat{H}}}$ to quantum circuit, as $ (\hat{H},r)  \gg U$, for every state $\psi$, let $\psi_1=\denote{U}(\psi)$, the distance of the circuit and the simulation is bounded by $\epsilon$, as $\dabs{\psi_1 -\denote{\eexp{\uapp{r}{\hat{H}}}}(\psi)}<\epsilon$. 
\end{theorem}

\begin{proof}
    Fully mechanized proofs are done in \rocq.
\end{proof}

\subsection{Canonicalization}\label{sec:compile-particlea}

After the particle transformation, we canonicalize the Hamiltonian program into Pauli-string form, where each Pauli string has a real amplitude.
We first define the canonicalized form below.

\begin{definition}[Pauli Canonical Form]\label{def:canon-forma}\rm 
Let $\pau \in \{I, X, Y, Z\}$, an $n$ qubit Hermitian Hamiltonian $\hat{H}$, is called Pauli Canonical $\sum_j r_{j} \sigma_j$, where $\sigma_j=\bigotimes_{k=0}^{n\sminus 1} \pau_{j,k}$ ($r_j$ is a real number).
\end{definition}

\Cref{def:canon-forma} states that every $n$-qubit program $\hat{H}$ can be viewed as a two-dimensional list. The outer list corresponds to a sum $\sum_j r_j \hat{P}_j$, and each inner element $\hat{P}_j$ is a fixed $n$ length tensor product $\bigotimes_{k=0}^{n-1} \mathpzc{P}_k$, representing a list of $n$ Pauli operations.
In \rocq, we implement the Pauli-string-based Hamiltonian and its intermediate Hamiltonian simulation formula as the following list structure.

{
\[
\begin{array}{l@{\quad}l@{\;\;}c@{\;\;}l}
\text{Hamiltonian} & \hat{H} & ::= & [(z_1,\sigma_1),\ldots,(z_m,\sigma_m)] \qquad z_j \in \mathbb{C} \\[0.2em]
\text{Real Hamiltonian} & \hat{H}_{\mathbb{R}} & ::= & [(r_1,\sigma_1),\ldots,(r_m,\sigma_m)] \qquad r_j \in \mathbb{R} \\[0.2em]
\text{Rotation sequence} & R & ::= & [(\theta_1,\sigma_1),\ldots,(\theta_m,\sigma_m)] \qquad \theta_j \in \mathbb{R}
\end{array}
\]
}

We now show the canonicalization theorem step-by-step, and we first show that every Hermitian Hamiltonian can be rewritten in a form with only real amplitudes.
\qsnd replaces each weight by its real part,
$\mathit{norm}([(z_j,\sigma_j)]_j) = [(\mathrm{Re}\,z_j,\sigma_j)]_j$.
The later stages need real weights, because $\exp(-i\theta\llbracket
\sigma \rrbracket)$ is unitary only when $\theta$ is real.

\begin{lemma}[Realness]\label{thm:realnessa}
If $\llbracket \hat{H} \rrbracket$ is Hermitian, then
$\llbracket \mathit{norm}(\hat{H}) \rrbracket = \llbracket \hat{H} \rrbracket$.
\end{lemma}

\begin{proof}
Each $\llbracket \sigma_j \rrbracket$ is Hermitian, as a tensor product
of Hermitian matrices, so
$\llbracket \hat{H} \rrbracket^\dagger = \sum_j \overline{z_j}\,\llbracket \sigma_j \rrbracket$.
Hence
$\llbracket \hat{H} \rrbracket
 = \tfrac12\bigl(\llbracket \hat{H} \rrbracket + \llbracket \hat{H} \rrbracket^\dagger\bigr)
 = \sum_j \mathrm{Re}(z_j)\,\llbracket \sigma_j \rrbracket
 = \llbracket \mathit{norm}(\hat{H}) \rrbracket$.
Mechanized in \texttt{QBlueHamiltonian.v}.
\end{proof}

The hypothesis is necessary: $[(i, Z)]$ denotes $iZ$, which is not
Hermitian, and $\mathit{norm}$ maps it to $[(0, Z)]$.

The second part removes the sequence operations.
Note that \Cref{def:canon-forma} does not need to have a sequence operation ($\circ$) because any sequence of single-qubit Paulis can be canonicalized to a singleton Pauli, described in \Cref{fig:pauli-compilea}. In fact, any valid Pauli-string-based Hamiltonian can be rewritten in the following form.

          \begin{wrapfigure}{r}{3.8cm}
$\begin{array}{|c|c|c|c|}
           \hline
             & X & Y & Z\\           \hline
           X & I & \iseq{i}{Z} & \iseq{-i}{Y} \\           \hline
           Y & \iseq{-i}{Z} & I & \iseq{i}{X} \\           \hline
           Z & \iseq{i}{Y} & \iseq{-i}{X} & I\\           \hline
           \end{array}$
            \caption{Pauli Merging}
            \label{fig:pauli-compilea}
          \end{wrapfigure}

A Hamiltonian is a list of Pauli strings with complex weights, and it
denotes the weighted sum of their matrices.  Strings may repeat in the
list.  We write $X(0)\circ Y(1)$ for the string $XY$, as in
\Cref{sec:overview}.  The input format is general because the $4^n$
Pauli strings on $n$ qubits form a basis for all $2^n \times 2^n$
matrices.  The simulation of $\hat{H}$ for time $r$ denotes
$\exp(-ir\llbracket \hat{H} \rrbracket)$, where
$\exp(M) = \sum_{k\ge 0} M^k/k!$; the series converges for every
matrix $M$, and $\exp(-irM)$ is unitary when $M$ is Hermitian.
Trotterization produces a rotation sequence, which denotes the product
of its rotations applied in list order.

Another key simplification is that a sequence of Pauli operations ($\pau \in \{I, X, Y, Z\}$) on a single site can be merged into a single Pauli operation, i.e., $\pau_1 \circ \pau_2$ can be rewritten as a single Pauli operation. Applying an identity $I$ with a $\pau$ results in $\pau$. 
The application results of other Pauli operations ($\pau_1$ and $\pau_2$) are defined in \Cref{fig:pauli-compilea}.
The two Pauli applications represent simplification: a sequence of Pauli group applications results in a single Pauli operation with a specific real-number amplitude.
We have the following lemma on canonicalization correctness.

\begin{lemma}[Canonical Compilation Correctness]\label{thm:canonical-gooda}\rm 
A $\hat{H}$ can be rewritten to a canonicalized form in \Cref{def:canon-forma}.
\end{lemma}

\begin{proof}
The proof relies on \Cref{thm:realnessa} and the sequence-operation merging above.
\end{proof}

\subsection{Trotterization}\label{sec:trottera}

The general workflow of the \qsnd Trotterization step works as follows. Given a time period $r$, and a desired error $\epsilon$, and an $n$-qubit Pauli-string-based Hamiltonian $\hat{H}=\sum_{j=0}^{d-1} \theta_j \hat{P}_j$ (with $\hat{P}_j=\bigotimes_{k=0}^{n-1} \mathpzc{P}_{j,k}$), we select a coefficient $\lambda$ such that $ \underbrace{\eexp{\frac{r}{\lambda}\theta_0\hat{P}_0} \circ ...\circ \eexp{\frac{r}{\lambda} \theta_j\hat{P}_j} \circ ...}_{N \ \cn{times}}$ approeximates $\eexp{{r \hat{H}}}$ to $\epsilon$.
The error is related to $r$ and $\theta_j$, detailed below.

The four Trotterization algorithms in \qsnd can be partitioned into two groups: the standard and QDrift \cite{Campbell_2019} groups.
The two standard algorithms essentially use the Lie-Trotter method. The two QDrift methods randomize the order of exponential operations, enabling faster Hamiltonian simulation.  We first state the general theorem based on an error bound $\epsilon$ (specialized to the two algorithms) and its proof in \rocq, then discuss it individually, unveiling the error bounds in different settings.

\begin{lemma}[Trotterization Correctness]\label{thm:trotter-gooda}\rm 
Given $\hat{H}=\sum_{j=0}^{d-1} \theta_j \hat{P}_j$, time $r$, and a desired error $\epsilon$, there is $\lambda$ and $N$, such that, $\eexp{\uapp{r}{\hat{H}}}$ is approximated by $\Pi^{N-1}_{k=0} U_k(d,\lambda,N,r)$, within the given error bound as: $\dabs{\denote{{\eexp{r \hat{H}}}} - \denote{{\Pi_{k=0}^{N-1} U_k(d,\lambda,N,r)}}}<\epsilon$.
\end{lemma}

\begin{proof}
The proof relies on four different algorithms. In \rocq, the proof is split into four different theorems for the four algorithms, mechanized in \code{QBlueTrotterProof.v}, \code{QBlueQdriftProof.v}, and \code{QBlueMarQSimProof.v}. These proofs proceed by induction on $N$, with key error-bound lemmas discussed below.
\end{proof}

As a running example, we use the following simple Ising model and present its Trotterized results, along with descriptions of the two algorithm groups.

\[
\hat{H}_{I1} = \sum_j r_h X(j) + \sum_{j} {Z(j)}\circ{Z(j\splus 1)}
    \tag{4.2.1}
  \label{eq:isinga}
\]

\subsubsection{The Standard Algorithms.}
We include first- and second-order standard Trotterization in our pipeline.
For first order Trotterization, we approximate $\eexp{\uapp{r}{\hat{H}}}$ by setting $\Pi^{N-1}_{k=0} U_k(d,\lambda,N,r)$ to be $\Pi^{d-1}_{k=0} \Pi^m \eexp{\frac{r}{m} \theta_j \hat{P}_j }$, i.e., for each Pauli string $\theta_j \hat{P}_j$, we construct an application sequence $\Pi^m \eexp{\frac{r}{m} \theta_j \hat{P}_j }$, where we repeat the simulation of $\theta_j \hat{P}_j$ for $m$ times with a scaled time slot $\frac{r}{m}$.
Each $U_k(d,\lambda,N,r)$ is defined as $\eexp{\frac{r}{m} \theta_j \hat{P}_j }$ with $N=d * m$ and $\lambda = m$.
For example, to simulate the Pauli-string-based Hamiltonian $\hat{H}_{I1}$ with a time slot $r$ and splitting number $m$, we compute the approximation of the exponent $\eexp{r{\hat{H}_{I1}}}$ as follows.

{\footnotesize
\begin{center}
$\hspace{-0.5em}
\begin{array}{l@{\;}c@{\;}l}
{\eexp{r{\hat{H}_{I1}}}}
&=& {{\Pi}^m\left(\eexp{\frac{r}{m}{(\sum_{j} \sapp{Z(j)}{Z(j\splus 1)} + r_h \sum_j X(j))}}\right)}\\
&=& {{\Pi}^m\left(\sapp{\sapp{\eexp{{{\frac{r}{m}}\sapp{Z(0)}{Z(1)}}}}{ \sapp{\eexp{{\frac{r}{m}}{\sapp{Z(1)}{Z(2)}}}}{...}}}{\sapp{\eexp{{\frac{r}{m}r_h}{X(0)}}}{\sapp{\eexp{{\frac{r}{m}r_h}{X(1)}}}{...}}} \right)}
\end{array}
$
\end{center}
}

The second-order Trotterization accounts for the relation between singleton Pauli strings. If we expand $\eexp{\uapp{r}{\hat{H}}}$ as a Taylor series, the second-order terms essentially pair-wise multiply singleton Pauli strings together. In the above example, other than the first-order terms, we also include the following terms, for all $j$:

{\small
\[
\frac{1}{2}\cdot X(j)\circ\sapp{Z(j)}{Z(j\splus 1)}\qquad \frac{1}{2}\cdot\sapp{Z(j)}{Z(j\splus 1)}\circ X(j)
\]
}

Including the two terms will improve the $m$ based on the same error bound, as now $m$ is related to $O(\frac{r^{1.5}}{\epsilon^{0.5}})$ instead of $O(\frac{t^2}{\epsilon})$ in the first order Trotterization.

\myparagraph{Proof Method for Error Bounds.}
Let $\hat{H}=\sum_{j=0}^{d\sminus 1} \theta_j \hat{P}_j$ with $N=d * m$, if we list all the Trotted Hamiltonian simulation sequences by listing all Pauli strings, we form an $N$-element set as $\hat{H}_k\in \{\frac{\theta_0}{m}\hat{P}_0,\frac{\theta_0}{m}\hat{P}_0,...,\frac{\theta_j}{m}\hat{P}_j,\frac{\theta_j}{m}\hat{P}_j,..., \frac{\theta_{d\sminus 1}}{m}\hat{P}_{d\sminus 1}\}$.
$\hat{H}$ can be viewed as a sum of $N$ different element Hamiltonians $\hat{H}_k$ in the set.
The error bound $\epsilon$ is related to $O(r^2)$, and we relate the error bound to different components as shown in the formula below (error bound is listed as $\epsilon'$).

{\small
\[
\displaystyle
\epsilon' = \displaystyle { \frac{r^2}{2} \sum_{k=1}^N \left\lVert 
(\sum_{j=k+1}^N \hat{H}_j) \circ \hat{H}_k - \hat{H}_k \circ (\sum_{j=k+1}^N \hat{H}_j)
\right\rVert}.
\label{eq:trottererr_stda}
\tag{4.2.2}
\]
}

The \qsnd compiler takes $\epsilon$ as input, and we find the corresponding $N$ by repeatedly applying the formula above until $\epsilon' < \epsilon$.

\subsubsection{The QDrift Group.}
QDrift is a randomized sampling method that samples subterms of the Hamiltonian according to their weights.
We manage the Pauli strings in $\hat{H}=\sum_{k=0}^{d\sminus 1} \theta_k \hat{P}_k$ as a set $\{\theta_0\hat{P}_0,..,\theta_{d\sminus 1}\hat{P}_{d\sminus 1}\}$. Let $\lambda = \sum_{k=0}^{d\sminus 1} \slen{\theta_k}$; we probabilistically pick a Pauli string in the set and repeat the process $N$ times; such picks are listed as a sequence $m_0, ...m_k,..., m_{N\sminus 1}$, each $m_k \in [0,d)$, and the probability of selecting a term $\theta_{k}\hat{P}_{k}$ is $|\theta_k|/\lambda$. The QDrift procedure produces an application sequence $\Pi_{k=0}^{N-1} \eexp{\frac{r\lambda}{N}\theta_{m_k}\hat{P}_{m_k} }$, i.e., each $U_k(d,\lambda,N,r)$ is set to $\eexp{\frac{r\lambda }{N}\theta_{m_k}\hat{P}_{m_k} }$ and $\eexp{\uapp{r}{\hat{H}}} \approx  \Pi_{k=0}^{N-1} U_k(d,\lambda,N,r)$.

We demonstrate the key difference between QDrift and standard Trotterization using the $\hat{H}_{I1}$ example above. To approximate the exponential $\eexp{r{\hat{H}_{I1}}}$, we randomly choose $N$ Pauli strings in the set $\{r_h X(0), r_h X(1),...,Z(0) \circ Z(1), Z(1) \circ Z(2), ...\}$, and then construct the Trottered sequence. One example choice is as follows.

{\footnotesize
\[\eexp{r{\hat{H}_{I1}}}
= \sapp{\sapp{\eexp{{\frac{r\lambda}{N}}{\sapp{Z(j)}{Z(j\splus 1)}}}}{\sapp{\eexp{{\frac{r\lambda}{N}}{\sapp{Z(k)}{Z(k\splus 1)}}}}{...}}}{\sapp{\eexp{{\frac{r\lambda}{N}r_h{X(k)}}}}{\sapp{\eexp{{\frac{r\lambda}{N}r_h}{X(0)}}}{...}}}
\]
}

MarQSim \cite{cao2024marqsimreconcilingdeterminismrandomness} improves on QDrift by noting that placing Pauli strings near each other increases the chance of canceling $\cn{CX}$ gates during circuit optimization when the program is compiled to digital circuits.
To increase the chance of grouping these relational Pauli strings, MarQSim uses a Markov chain graph with nodes corresponding to different Pauli strings, increasing the likelihood of selecting them.
Our evaluation shows that MarQSim produces an optimized circuit compared to QDrift when we perform digital circuit compilation.
However, the improvement becomes slim as the Hamiltonian size increases, and MarQSim fails for large Hamiltonians ($> 10K$ Pauli strings) because sampling its Markov chain becomes difficult. It is also questionable for handling analog-based compilation.

\myparagraph{Proof Method for Error Bounds.}
The error rate $\epsilon$ in QDrift is $O(\frac{r^2}{N})$, where $N$ depends on the input $\epsilon$ as follows. 
{\small
\[
\displaystyle
N = \displaystyle{ \frac{2\lambda^2 r^2}{\epsilon}}
\tag{4.2.3}
\label{eq:trottererr_qdrifta}
\]
}

The \rocq proof of the QDrift correctness relies on a key error bound lemma based on the error bound in \Cref{eq:trottererr_qdrifta}.

\subsection{Synthesization}\label{sec:decomposea}

Synthesization applies the application sequence of matrix exponentials of Pauli strings generated from Trotterization to quantum circuits, either digital or analog.
To synthesize an application sequence $\Pi_{j} \eexp{\theta_j\hat{P}_j}$, it suffices to describe the synthesization of a single matrix exponential term $\eexp{\theta_j \hat{P}_j}$. We discuss the digital and analog-based synthesization below.

\begin{algorithm}[t]
{\small
\caption{Digital Synthesization of \( U = \exp(-i \theta \hat{P}) \) for a Pauli string \(\hat{P}\)}
\label{fig:digitaldecoa}
\begin{algorithmic}[1]
\STATE \textbf{Input:} Pauli string \( \hat{P} = \bigotimes_{k=0}^{n\sminus 1} \mathpzc{P}_{k} \), time period \( \theta \)
\STATE \textbf{Output:} Digital circuit \( U = \exp(-i \theta \hat{P}) \)
\STATE

\STATE Collect sites \( \{m_1, m_2, \ldots, m_l\} \) where \( \mathpzc{P}_{m_j} \neq I \), with \( l \leq n \)

\STATE \( U \leftarrow \cn{Rz}(2\theta)(m_1) \)
\STATE $j = m_1$
\FOR{ \( m_j \in \{m_2, \ldots, m_l\} \)}
    \STATE \( U \leftarrow \cn{CX}(m_{j-1}, m_j) \circ U \circ \cn{CX}(m_{j-1}, m_j) \)
\ENDFOR

\FOR{ \( m_j \in \{m_1, \ldots, m_l\} \)}
    \IF{\( \mathpzc{P}_{m_j} = Y \)}
        \STATE \( U \leftarrow \cn{Rz}\left(\tfrac{\pi}{2}\right)(m_j) \circ \cn{H}(m_j) \circ U \circ \cn{H}(m_j) 
        \circ \cn{Rz}\left(-\tfrac{\pi}{2}\right)(m_j) \)
    \ELSIF{\( \mathpzc{P}_{m_j} = X \)}
        \STATE \( U \leftarrow \cn{H}(m_j) \circ U \circ \cn{H}(m_j) \)
    \ENDIF
\ENDFOR

\STATE \textbf{Return} \( U \)
\end{algorithmic}
}
\end{algorithm}

\myparagraph{Digital Synthesization.}
The digital synthesization synthesizes the simulation of a Pauli string, $\eexp{\theta_j \hat{P}_j}$, to elementary quantum gates, e.g., single-qubit gates, such as \cn{H}, \cn{Rz}, \cn{Ry}, and \cn{Rx} gates, as well as a two-qubit controlled-NOT gate \cn{CX} ($\cn{CX}(x,y)$ controlling on $x$ and applying $\cn{X}$ gate on $y$).

The procedure focuses on the non-$I$ terms in a Pauli string $\hat{P}=\bigotimes_{j=0}^{n-1} \mathpzc{P}_j$. If $\hat{P}$ has only one non-$I$ term, the compilation generates a rotation gate, listed in \Cref{fig:paulisima}.
In the Bloch qubit sphere interpretation \cite{mike-and-ike}, the rotation gates $\cn{Rx}$, $\cn{Ry}$, $\cn{Rz}$ can be expressed as some Hamiltonian simulations of Pauli matrices ($\cn{Rz}(2 \theta)(j)$ applies a $Z$-axis rotation in the $j$-th qubit), which essentially suggests that the simulation of singleton Pauli $X$, $Y$, and $Z$ terms are the above phase rotation gates.
For example, the gate synthesis of $\eexp{{\frac{r}{n}}{X(j)}}$ generates an $\cn{Rx}$ gate, as $\cn{Rx}(\frac{2r}{n})(j)$. 
We describe the algorithm in \Cref{fig:paulisima} for a Pauli string with more than one non-$I$ term, i.e., a $k$-local Pauli string with $k>1$.

\begin{figure}[t]
{\small
\begin{center}
$
\begin{array}{c}
{\eexp{{\theta}{X}}}=\cn{RX}(2{\theta})
=
\begin{psmallmatrix}
\cos{{\theta}} & -i\sin{{\theta}}\\
-i\sin{{\theta}} & \cos{{\theta}}
\end{psmallmatrix}
\qquad
{\eexp{{\theta}{Y}}}=\cn{RY}(2{\theta})
=
\begin{psmallmatrix}
\cos{{\theta}} & -\sin{{\theta}}\\
\sin{{\theta}} & \cos{{\theta}}
\end{psmallmatrix}
\\[0.3em]
{\eexp{{\theta}{Z}}}=\cn{RZ}(2{\theta})
=
\begin{psmallmatrix}
\eexp{{\theta}} & 0\\
0 & \cn{exp}{(i {\theta})}
\end{psmallmatrix}
\end{array}
$
\end{center}
}
  \caption{Pauli matrices simulated to rotation gates. $\eexp{\uapp{\theta}{\pau}}$ can be thought as simulating $\pau$ on time $\theta$.}
  \label{fig:paulisima}
\end{figure}

\begin{figure*}[t]
{\small
\begin{tabular}{c@{$\;\;=\;\;$}c}
\begin{minipage}{0.22\textwidth}
$
\eexp{\theta{\sapp{Z(j)}{Z(j\splus 1)}}}
$
\end{minipage}
&
\begin{minipage}{0.3\textwidth}
{
  \Qcircuit @C=0.5em @R=0.5em {
    \lstick{} & & \ctrl{1}       & \qw  & \ctrl{1}      &\qw\\
    \lstick{} & & \targ         & \gate{\cn{Rz}(2\theta)}       & \targ    & \qw 
    }
}   
\end{minipage}
\end{tabular}
\hspace{1em}
\begin{minipage}{0.48\textwidth}
{\small
$
\eexp{\theta ZZ}=
\begin{psmallmatrix}
\cn{exp}(\sminus i\frac{\theta}{2}) & 0 & 0 & 0\\
 0 & \cn{exp}(i\frac{\theta}{2}) & 0 & 0\\
 0 & 0 & \cn{exp}(i\frac{\theta}{2}) & 0\\
 0 & 0 & 0 & \cn{exp}(\sminus i\frac{\theta}{2})
\end{psmallmatrix}
$
}
  \end{minipage}
}
\vspace*{-0.8em}
\caption{Circuit and unitary for $ZZ$ Interaction. The left circuit applies on $j$ and $j\splus 1$ qubits.}
\label{fig:zz-intera}
\end{figure*}

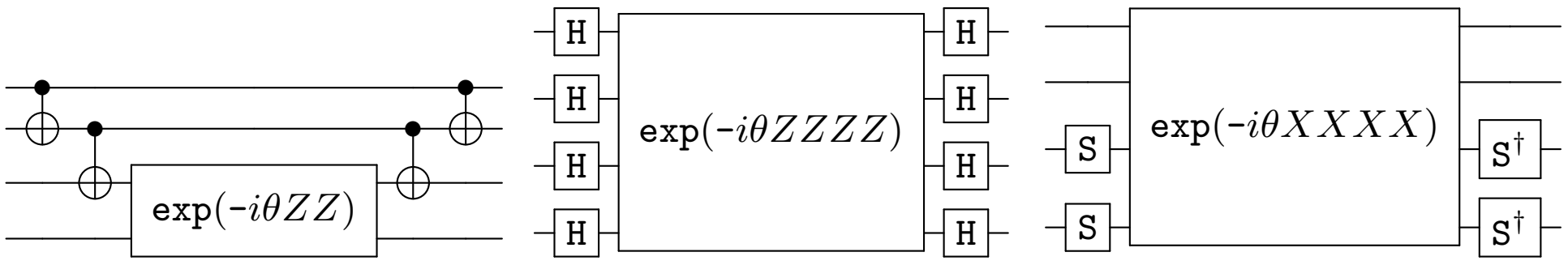
\begin{figure*}[t]
{\small
\begin{minipage}[b]{0.3\textwidth}
{\centering
{
  \Qcircuit @C=0.5em @R=0.5em {
    \lstick{} & \ctrl{1} & \qw          & \qw                            & \qw      & \ctrl{1} & \qw      \\
    \lstick{} & \targ    & \ctrl{1}     & \qw                            & \ctrl{1} & \targ & \qw \\
    \lstick{} & \qw      & \targ        & \multigate{1}{\eexp{\theta ZZ}} & \targ    & \qw   & \qw \\
    \lstick{} & \qw      & \qw          & \ghost{\eexp{\theta ZZ}}      & \qw      & \qw   & \qw
    }
}}
\subcaption{$\eexp{{\theta}{ZZZZ}}$ gate.}
\label{fig:zzzz-gatea}  
\end{minipage}
\begin{minipage}[b]{0.29\textwidth}
{\centering
{
  \Qcircuit @C=0.5em @R=0.5em {
    \lstick{} & \gate{\cn{H}} & \multigate{3}{\eexp{{\theta}{ZZZZ}}} & \gate{\cn{H}} & \qw   \\
    \lstick{} &\gate{\cn{H}}  & \ghost{\eexp{{\theta}{ZZZZ}}}        & \gate{\cn{H}} & \qw  \\
    \lstick{} & \gate{\cn{H}} & \ghost{\eexp{{\theta}{ZZZZ}}}        & \gate{\cn{H}} & \qw  \\
    \lstick{} & \gate{\cn{H}} & \ghost{\eexp{{\theta}{ZZZZ}}}        & \gate{\cn{H}} & \qw
    }
}}
\subcaption{$\eexp{{\theta}{XXXX}}$ gate.}
\label{fig:xxxx-gatea}  
\end{minipage}
\begin{minipage}[b]{0.29\textwidth}
{\centering
{
  \Qcircuit @C=0.5em @R=0.5em {
    \lstick{} & \qw & \multigate{3}{\eexp{{\theta}{XXXX}}} & \qw & \qw   \\
    \lstick{} &\qw  & \ghost{\eexp{{\theta}{XXXX}}}        & \qw & \qw  \\
    \lstick{} & \gate{\cn{S}} & \ghost{\eexp{{\theta}{XXXX}}}        & \gate{\sdag{\cn{S}}} & \qw  \\
    \lstick{} & \gate{\cn{S}} & \ghost{\eexp{{\theta}{XXXX}}}        & \gate{\sdag{\cn{S}}} & \qw
    }
}}
\subcaption{$\eexp{{\theta}{XXYY}}$ gate.}
\label{fig:xxyy-gatea}  
\end{minipage}
}
\caption{Synthesized circuits for some long Pauli strings.}
\label{fig:bosongatesa}
\end{figure*}

We first show the gate synthesization of the simple Ising model $\hat{H}_{I1}$ in \Cref{eq:isinga}, which produces a sequence of $\eexp{\frac{r}{n} {Z(j)}\circ{Z(j\splus 1)}}$ and $\eexp{\frac{r}{n} r_h X(j)}$ operations (the sequence length is related to $n$). If $n$ is large enough, the operation sequence is a good approximation to $\eexp{r \hat{H}_{I1}}$.
The $\eexp{\frac{r}{n} r_h X(j)}$ operation produces a \cn{Rx} gate \cite{mike-and-ike}, while $\eexp{ \frac{r}{n} {Z(j)}\circ{Z(j\splus 1)}}$ produces a digital circuit, performing two controlled-not gates and an $\cn{Rz}(2\frac{r}{n})$ gate in the middle for the $j$-th and $j\splus 1$-th qubits,
i.e., $\cn{Rz}(2\theta)$ is an $z$-axis phase rotation gate.

The above assumes a two-local Pauli string with non-$I$ terms being $Z$.
We show an example of a long Pauli string synthesization in \Cref{fig:bosongatesa}.
After Trotterizing $\hat{H}$, we generate a sequence of Hamiltonian simulations of Pauli strings.
Some possible Pauli string terms might be $4$-local, as $Z \otimes Z \otimes Z \otimes Z$ (abbreviated as $ZZZZ$), $X\otimes X \otimes X \otimes X$ (abbreviated as $XXXX$) and $X\otimes X \otimes Y \otimes Y$ (abbreviated as $XXYY$).
The simulation of the $ZZZZ$ term yields a quadruple-$Z$ interaction in \Cref{fig:zzzz-gatea}, with a $ZZ$ interaction applied in the middle, as shown in \Cref{fig:zz-intera}, surrounded by controlled-NOT gates.
The digital gate synthesization of $\eexp{{\theta}{XXXX}}$ and $\eexp{{\theta}{XXYY}}$ is based on the quadruple $Z$ interaction with repairs for $X$ and $Y$ conversion, which results in the circuits in \Cref{fig:xxxx-gatea,fig:xxyy-gatea}.
We show the correctness lemma below.

\begin{lemma}[Ditigal Synthesization Correctness]\label{thm:decompose-gooda}\rm 
Given $\hat{P}$, $\eexp{\theta \hat{P}}$ is correctly synthesized by \Cref{fig:digitaldecoa}, which is based on the elementary gates, $\{\cn{H}, \cn{Rx}, \cn{Ry}, \cn{Rz}, \cn{CX}\}$.
\end{lemma}

\begin{proof}
    We proved these fully mechanized proofs by induction on the Pauli canonical forms and mechanized them in \code{QBlueSynthDigitalProof.v}. By applying \Cref{fig:digitaldecoa}, we produce a unitary $U$ for $\eexp{\theta \hat{P}}$. We then interpret the two terms as matrix representations in VOQC \cite{VOQC} and equate them.
\end{proof}

\begin{algorithm}[h]
{\small
\caption{IBM-based Analog Synthesization of \( U = \exp(-i \theta \hat{P}) \) for $2$-Local Pauli Strings}
\label{fig:analogdecoa}
\begin{algorithmic}[1]
\STATE \textbf{Input:} $2$-local Pauli string \( \hat{P} = \bigotimes_{k=0}^{n\sminus 1} \mathpzc{P}_{k} \), time period \( \theta \)
\STATE \textbf{Output:} Analog Hamiltonian simulation \( U = \Pi^m\eexp{  r \hat{P}} \)
\STATE
\STATE Identify non-identity sites \( m_1 \) and \( m_2 \) where \( \mathpzc{P}_{m_j} \neq I \)

\STATE \( U \leftarrow \eexp{ \theta Z({m_1}) \circ Z({m_2})} \)

\FOR{ \( m_j \in \{m_1, m_2\} \) }
    \IF{\( \pau_{m_j} \neq Z \)}
            \STATE \textbf{let} $U_1 =  \eexp{\frac{\pi}{4} X(m_j)} \circ  \eexp{\frac{\pi}{4} Z(m_j)} \circ  \eexp{\frac{\pi}{4} X(m_j)}$ \textbf{in}
                 \STATE  \( U \leftarrow  U_1 \circ U \circ U_1\) 
    \ENDIF
    \IF{\( \pau_{m_j} = Y \)}
       \STATE \( U \leftarrow \eexp{ \tfrac{\pi}{4} Z({m_j})} \circ U \circ \eexp{\tfrac{7\pi}{4} Z({m_j})} \)
    \ENDIF
\ENDFOR

\STATE \textbf{Return} \( U \)
\end{algorithmic}
}
\end{algorithm}

\begin{algorithm}[h]
{\small
\caption{Indiana Analog Synthesization of \( U = \exp(-i \theta \hat{P}) \)}
\label{fig:incoa}
\begin{algorithmic}[1]
\STATE \textbf{Input:} Pauli string \( \hat{P} = \bigotimes_{k=0}^{n\sminus 1} \mathpzc{P}_{k} \), time period \( \theta \)
\STATE \textbf{Output:} Analog circuit \( U = \exp(-i \theta \hat{P}) \)
\STATE

\STATE Collect sites \( \{m_1, m_2, \ldots, m_k\} \) where \( \mathpzc{P}_{m_j} \neq I \), with \( l \leq n \)

\STATE \( \hat{P'} = I \)
\FOR{ \( m_j \in \{m_1, \ldots, m_k\} \)}
    \STATE \( \hat{P'} \leftarrow \hat{P'} \circ X(m_j) \)
\ENDFOR

\STATE  $U \leftarrow \eexp{\uapp{\theta}{\hat{P'} }}$

\FOR{ \( m_j \in \{m_1, \ldots, m_k\} \)}
    \IF{\( \mathpzc{P}_{m_j} = Y \)}
        \STATE \( U \leftarrow \eexp{\frac{\pi}{4} Z(m_j)} \circ U \circ \eexp{\frac{7\pi}{4} Z(m_j)} \)
    \ELSIF{\( \mathpzc{P}_{m_j} = Z \)}
        \STATE \textbf{let} $U_1 =  \eexp{\frac{\pi}{4} X(m_j)} \circ  \eexp{\frac{\pi}{4} Z(m_j)} \circ  \eexp{\frac{\pi}{4} X(m_j)}$ \textbf{in}
                 \STATE  \( U \leftarrow  U_1 \circ U \circ U_1\) 
    \ENDIF
\ENDFOR

\STATE \textbf{Return} \( U \)
\end{algorithmic}
}
\end{algorithm}

\myparagraph{Analog Synthesization.}\label{sec:ana-syna}
Many quantum computers provide backends for analog quantum computation; that is, we can view a quantum computer as a Hamiltonian simulator that runs small Hamiltonians sequentially. Each simulation step in a quantum simulator takes a positive real number $\theta$ and a Pauli string $\hat{P}$ and produces the simulation result $\eexp{\theta \hat{P}}$. The procedure is machine-dependent. 
For example, an IBM machine can simulate one- or two-local Pauli strings in the set $\{X, Z, ZZ\}$ ($ZZ$ denotes a two-local Pauli string with only $Z$ Pauli operations). Another machine (Indiana), developed by Richerme \emph{et al.} \cite{richerme2025multimodeglobaldrivingtrapped,Kyprianidis_2024}, permits the simulation of $k$-local Pauli strings in the set $\{Z, X^*\}$, where $X^*$ permits a Pauli string simulation with an arbitrary number of $X$ Paulis.
We show two analog synthesization algorithms in \Cref{fig:analogdecoa,fig:incoa} to compile our Pauli strings to the IBM and Indiana machines, respectively.

To perform analog synthesization for the IBM machine, we assume that a two-local Hamiltonian is achieved by applying a perturbative gadget.
We can simulate a one-local Pauli using the matrix exponential formulas in \Cref{fig:paulisima}.
The Indiana machine allows simulating a $k$-local Pauli string.

As an example, to simulate a simple Ising system $\hat{H}_{I1}$ above, we can directly perform the simulation $\eexp{\frac{r}{n}\sapp{Z(j)}{Z(k)}}$ and $\eexp{\frac{r}{n}r_h {X(j)}}$ for sites $j$ and $k$, in the IBM setting. If $\frac{r}{n}$ is negative, we can instead simulate a positive number $\frac{r}{n}+2m\pi$. Direct Hamiltonian simulation enables more efficient circuit compilation.
The analog synthesization step above implements the $ZZ$-interaction gate in \Cref{fig:zz-intera} and explains why some interaction gates are much more efficient than elementary gates on IBM machines, as shown in Peng \emph{et al.} \cite{10.1145/3632923}.

To perform the simulation $\eexp{{\theta}{XXYY}}$ in the Indiana machine, we can construct the simulation procedures (\Cref{fig:incoa}) as follows:
We construct the simulation operation $\eexp{\theta XXXX}$, with the single qubit simulations $\eexp{\frac{\pi}{4} Z}$ and $\pexp{\frac{7\pi}{4} Z}$ around the last two qubits.

Other quantum machines may permit different Pauli string sets, such as Quera machines \cite{quera2024}. Consequently, the machine Hamiltonian analog synthesization step must employ different analog synthesization methods for each quantum machine.
We show the correctness lemma for the two machines below.

\begin{lemma}[Analog Synthesization Correctness]\label{thm:analog-decompose-goodacrefa}\rm 
Given $\hat{P}$, $\eexp{\theta \hat{P}}$ is correctly synthesized by \Cref{fig:analogdecoa,fig:incoa}, based on the simulations of Pauli strings permitted in the IBM machine ($\{X, Z, ZZ\}$) and the Indiana machine ($\{Z, X^*\}$), respectively.
\end{lemma}

\begin{proof}
    We mechanized the proofs using the same methodology as in \Cref{thm:decompose-gooda}. Here, we equate the two matrix interpretations of $\eexp{\theta \hat{P}}$ and the unitaries generated via \Cref{fig:analogdecoa,fig:incoa}.  The proofs are mechanized in \code{QBlueSynthProof.v}.
\end{proof}

 \end{document}